\documentclass[traditabstract]{aa} 

\usepackage{graphicx}
\usepackage{txfonts}
\usepackage{natbib}
\usepackage{afterpage}

\bibpunct{(}{)}{;}{a}{}{,}

\renewcommand{\deg}{^\circ}
\newcommand{\ergs}{\mathrm{erg/s}}
\newcommand{\Hz}{\mathrm{Hz}}
\newcommand{\Kel}{\mathrm{K}}
\newcommand{\micron}{\rm \mu m}
\newcommand{\pc}{\mathrm{pc}}
\newcommand{\Spitzer}{\textit{Spitzer}}
\newcommand{\dd}{\mathrm{d}}
\newcommand{\amax}{a_\mathrm{max}}
\newcommand{\amin}{a_\mathrm{min}}
\newcommand{\Fnu}{F_{\nu}}
\newcommand{\fDH}{f_\mathrm{DH}}
\newcommand{\fIH}{f_\mathrm{IH}}
\newcommand{\Lnu}{L_{\nu}}
\newcommand{\Lbol}{L_\mathrm{bol}}

\newcommand{\nuFnu}{\nu F_{\nu}}

\newcommand{\No}{N_{0}}
\newcommand{\Ntot}{N_\mathrm{tot}}

\newcommand{\PhiV}{\Phi_{V}}
\newcommand{\PhiVo}{\Phi_{V;0}}
\newcommand{\Pfoc}{P_{0}(r,z)}
\newcommand{\Rcl}{R_\mathrm{cl}}
\newcommand{\Rclo}{R_{\mathrm{cl};0}}
\newcommand{\Rout}{R_\mathrm{out}}
\newcommand{\rrev}{r_{\tau_K}}
\newcommand{\rsub}{r_\mathrm{sub}}
\newcommand{\sigs}{\sigma_\mathrm{s}}
\newcommand{\tauV}{\tau_{V}}
\newcommand{\tausi}{\tau_\mathrm{Si}}
\newcommand{\taucl}{\tau_\mathrm{cl}}
\newcommand{\Tsub}{T_\mathrm{sub}}

\begin{document}
   \title{The dusty heart of nearby active galaxies}

   \subtitle{II. From clumpy torus models to physical properties of dust around AGN }

\titlerunning{The dusty heart of nearby active galaxies. II.}

\author{S.~F.~H\"onig\inst{1,2} \and
M.~Kishimoto\inst{2}
}

\offprints{S.~F. H\"onig \\ \email{shoenig@physics.ucsb.edu}}

\institute{
University of California in Santa Barbara, Department of Physics, Broida Hall, Santa Barbara, CA 93109, USA  \and
Max-Planck-Institut f\"ur Radioastronomie, Auf dem H\"ugel 69, 53121 Bonn, Germany
}

   \date{Received June 10, 2009; accepted August 18, 2010}

 
\abstract{
With the possibilities of high spatial resolution imaging and spectroscopy as well as infrared (IR) interferometry, the dusty environments (= ``dusty torus'') of active galactic nuclei (AGN) are now in reach of observations. Following our paper I on ground-based mid-IR spectro-photometry (H\"onig et al. 2010), we present an upgrade to our radiative transfer model of 3-dimensional clumpy dust tori. The upgrade with respect to H\"onig et al. (2006) concerns an improved handling of the diffuse radiation field in the torus which is approximated by a statistical approach. The models are presented as tools to translate classical and interferometric observations into characteristic properties of the dust distribution. We compare model SEDs for different chemical and grain-size compositions of the dust and find that clouds with standard ISM dust and optical depth $\tauV\sim50$ appear in overall agreement with observed IR SEDs. By studying parameter dependencies, it is shown that type 1 AGN SEDs, in particular the mid-IR spectral index, can be used to constrain the radial dust cloud distribution power-law index $a$, while other parameters are more difficult to assess using SEDs only. Interferometry adds important additional information for modeling when interpreted simultaneously with the SED. Although type 2 AGN can, in principle, be used to constrain model parameters as well, obscuration effects make the analysis more ambiguous. We propose a simple, interferometry-based method to distinguish between ``compact'' and ``extended'' radial dust distributions without detailed modeling of the data and introduce a way to easily determine individual or sample average model parameters using the observed optical depth in the silicate feature and the mid-IR spectral index.

}
   \maketitle
%

\section{Introduction}\label{sec:intro}

The Unification Scheme of active galactic nuclei (AGN) is relying on a toroidal region filled with molecular gas and dust to explain the observed dichotomy of broad- and narrow-line AGN \citep{Ant93,Urr95}. The presence of this optically-thick region is well supported by direct and indirect observational evidence. \citet{Mil83} showed that the narrow emission-line galaxy NGC~1068 shows broad optical emission lines in polarized light, scattered outward from the innermost region which is obscured by the ``dust torus''. Several succeeding studies confirmed the presence of a hidden broad-line AGN in many narrow-line objects \citep[e.g.][and references therein]{Mor00}. From number statistics of obscured type-2 AGN and non-obscured type-1 AGN as well as from direct observations of the pc-scaled molecular gas in nearby AGN, it is inferred that the torus is not only optically thick but also geometrically thick, with a characteristic scale height $h/r \sim 1$ \citep[e.g.][]{Mai95,Lac04,Mar06,Hic09}.

Obscuration of the torus does not only affect the optical broad lines but can also be observed in X-ray and optical/UV continuum emission from the accretion disk and its immediate vicinity. Optically-obscured type-2 AGN are usually associated with high Hydrogen column densities in the X-ray, from $\sim10^{23}\,{\rm cm}^{-2}$ to Compton-thick columns of $>10^{24-25}\,{\rm cm}^{-2}$ \citep[e.g.][]{Shi06}. The material causing these high column densities is generally believed to reside in the dust torus, tracing the gaseous component which dominates the total mass of the torus. Moreover, strong reflection on parsec scales, as observed in the X-ray continuum around 30\,keV and associated with large equivalent width of the Fe-K$\alpha$ line, are interpreted as signs for the presence of the torus.

Since a sizable portion of the torus consists of dust, the UV/optical accretion disk emission heats up the dust which thermally re-radiates the received energy in the infrared at temperatures of some 100\,K up to the dust sublimation temperature of $\sim$1500\,K. These temperatures are reached on scales of about 0.1 to 100\,pc, depending on the actual AGN luminosity, torus geometry, and dust composition. The thermal radiation from the torus dominates the total near- to mid-infrared (IR) emission of the AGN (``red bump'') with a noticeable cut-off at around 1\,$\micron$ where the UV/optical accretion disk emission starts to dominate (``big blue bump''). At around 10 and 18\,$\micron$, type-1 AGN show some broad spectral features caused by hot silicate dust which is believed to be associated with the inner part of the torus. On the other hand, these silicate features appear in absorption in most type 2 AGN where only cooler dust is seen.

Due to the small spatial scales at which the torus resides, it is a difficult task to directly resolve it by infrared observations. Thus, first reports of successful resolution of the nucleus of an AGN involved interferometric techniques. \citet{Wit98} used bispectrum speckle interferometry to resolve the nucleus of the Seyfert 2 galaxy NGC~1068 in the $K$-band, followed-up by additional $H$-band observations \citep{Wei04}. \citet{Swa03} presented $K$-band long-baseline interferometry of the type-1 AGN NGC~4151 which arguably resolve the innermost hot region of the dust torus \citep[see][for this interpretation]{Kis07}. More recently, VLTI/MIDI long-baseline mid-IR spectro-interferometry directly revealed the parsec-scaled dust emission sizes in the wavelength band from 8 to 13\,$\micron$ for a number of nearby type 1 and type 2 AGN \citep{Jaf04,Tri07,Bec08,Rab09,Tri09}. These observations finally confirmed the basic picture of the dust torus while more detailed characteristics remain unclear.

The dust torus has been the subject of several kinds of models in order to extract physical properties from broad SED and, in some cases, interferometric observations. Initially, most authors used smooth dust distributions with different kind of radial and vertical density profiles \citep[e.g.][]{Pie93,Gra94,Efs95,Sch05,Fri06}. It was, however, early noted that the dust is most probably arranged in clouds instead of being smoothly distributed \citep[e.g.][]{Kro88,Tac94,Hon07b}. This idea received further support by interferometric observations which arguably rule out smooth dust distributions \citep{Jaf04,Tri07}. Several radiative transfer models have been developed to account for 2D or 3D clumpy dust distributions \citep{Nen02,Dul05,Hon06,Sch08}. All of these models appear in more or less good agreement with observations, while the resulting torus or cloud properties differ significantly. \citet{Hon06} used optically thick dust clouds and a low torus volume filling factor to simultaneously model near- and mid-IR photometry and interferometry of NGC~1068 \cite[see also][]{Hon07a,Hon08b}, while \citet{Sch08} model a torus with high volume filling factor and optically thin clouds for a similar set of observations of the Circinus galaxy. Although the actual properties of the dust clouds are not yet constrained by observations, theoretical predictions and hydrodynamic simulations arguably favor small and compact optically-thick clouds \citep[e.g.][]{Vol04,Bec04,Hon07b,Sch09}.

In this paper, we present an update of our radiative transfer model of 3D clumpy AGN tori presented in \citet{Hon06}. In particular, a better handling of the diffuse radiation field inside the torus has been implemented to overcome the limitations discussed in the Appendix A.2 of \citet{Hon06}, and the possibility of different dust compositions and grain sizes is now included. The improved handling of the diffuse radiation field has been proposed in \citet{Hon08phd} and is described in more detail in Sect.~\ref{sec:dirindir}. We intend to provide models for observers and come up with some simple correlations between observational and model parameters. Some emphasis will be put on the radial distribution of the dust and dust grain size and chemical composition, the latter being only marginally explored in literature despite observational evidence that the grain composition might be crucial \citep[e.g.][]{Kis07,Kis09a}.

In Sect.~\ref{sec:torfun} we outline our model strategy and methods and introduce the mathematical background of the clumpy torus model including model parameters. Following in Sect.~\ref{sec:emcld} simulation results of individual dust clouds are presented. These simulations build up the dust cloud databases which serve as an input to the torus model simulations. In Sect.~\ref{sec:resdis} we show our results from the torus simulations and discuss various aspects of interpreting SEDs and IR interferometry. The results are summarized in Sect.~\ref{sec:sumconc}.

\section{Torus model fundamentals}\label{sec:torfun}

In this section, we will describe the mathematical and strategic base of our torus model and discuss underlying assumptions. First we outline the general model strategy and compare it conceptually to torus models in literature. Then, we describe the theoretical basis and parameters involved in the model.

\subsection{Model strategy}\label{sec:strategy}

The most straight-forward way of modeling clumpy dust tori in 3 dimensions is direct Monte Carlo simulation of a geometrically well-defined, statistically arbitrary distribution of dust clouds spread on a model grid, as demonstrated by \citet{Sch08}. There are, however, several practical problems that occur. When solving the radiative transfer equation by Monte Carlo simulations, it is important to properly sample optically thick surface regions by enough grid cells so that each cell is optically thin. Otherwise, emission temperatures will be underestimated leading to a wrong final source function which affects torus SEDs and images. In principle, adaptive grids can be used, but this may become a difficult task when aiming for $>10^3-10^4$ randomly-arranged clouds. Thus, the number of model clouds has to be small, the total optical depth of each cloud has to be limited, and/or the volume filling factors have to be rather large, of the order of $\PhiV\sim1$ \citep[see also][]{Dul05}. \citet{Sch08} showed that the SED of the nucleus of the Circinus galaxy as well as the position-angle- and baseline-dependence of the visibility is in good agreement with this kind of Monte Carlo models. Still, such direct Monte Carlo simulations take a lot of time and are, thus, not very flexible for modeling of observations.

As mentioned in Sect.~\ref{sec:intro}, it is most likely that individual torus clouds are actually optically thick (see also Sect.~\ref{sec:cloud_prop}). 
Based on optically thick clouds and the assumption of a low volume filling factor, $\PhiV \ll 1$, \citet{Nen02} used a probabilistic approach for cloud heating and obscuration, depending on several model parameters \citep[see also][]{Nat84,Nen08a}. \citet{Nen08b} presented model SEDs simulated via this approach which are in general agreement with observations. \citet{Bec05} used a similar probabilistic model to reproduce the high spatial resolution SED of NGC~1068. The benefit of this modeling approach is its time-efficient calculation of average model SEDs and average brightness distributions. On the other hand, as noted by \citet{Nen08b}, the probabilistic method is not capable of predicting variations in the overall SED and brightness distribution or small-scale spatial and position-angle variations (=''clumpiness variations'') of interferometric visibilities and phases caused by the random distribution of dust clouds in the torus.

\begin{figure}
\centering
\includegraphics[width=0.45\textwidth]{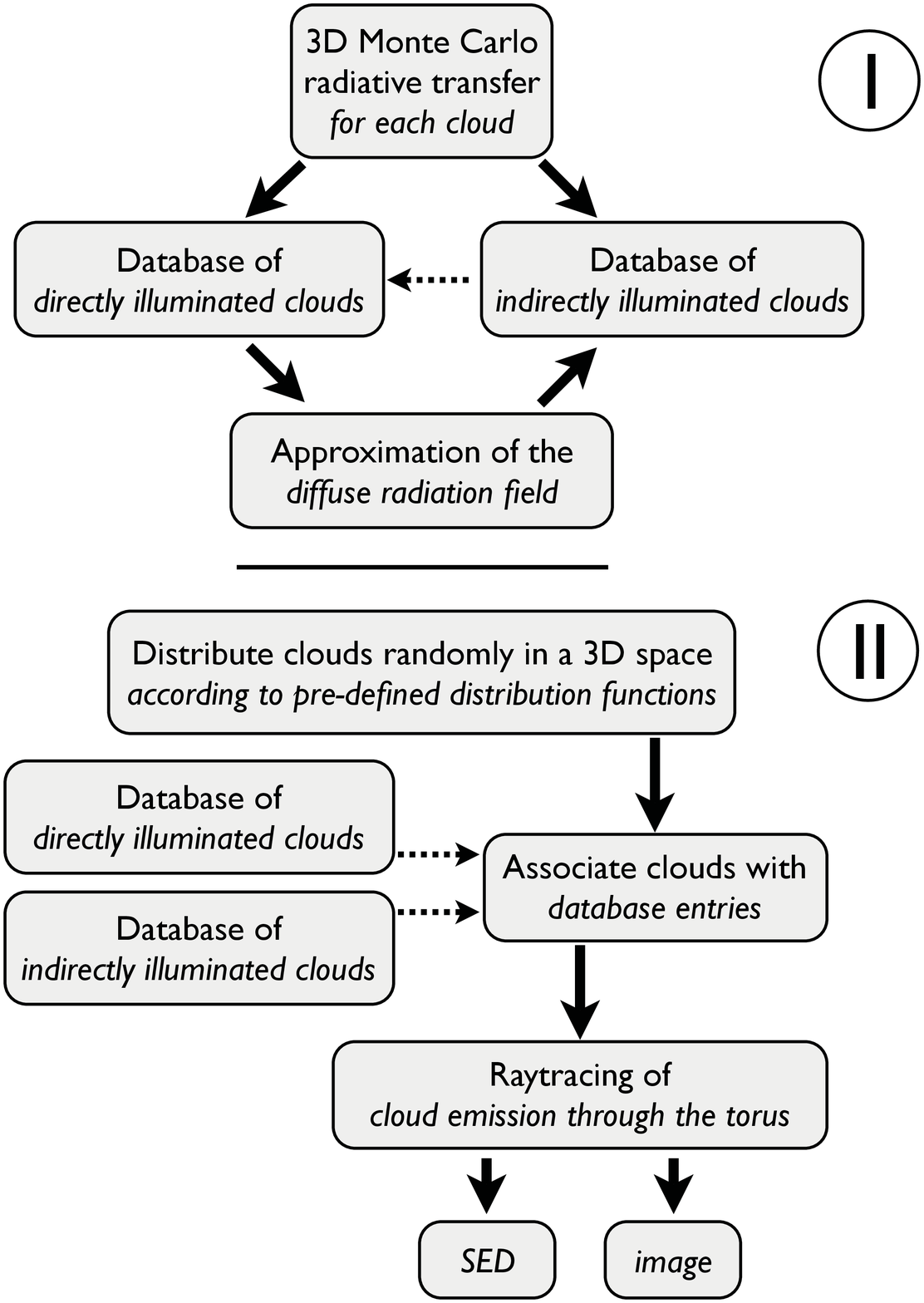}
\caption{Flow chart of our method for radiative transfer simulations of 3D clumpy tori. The strategy follows \citet{Nen08a}, however resolving the need for a probabilistic treatment of the dust distribution. In a first step, we simulate databases of directly- (see Sect.~\ref{sec:cloud_prop}) and indirectly-illuminated clouds (see Sect.~\ref{sec:indheat}). The step using the indirectly-illuminated clouds as a heating source for directly-illuminated clouds (dashed arrow) would close an iterative loop which we argue in Sect.~\ref{sec:dirindir} converges satisfactorily after the initial loop. In the second step, clouds are randomly distributed in a torus according to the model parameters (see Sect.~\ref{sec:modparams}). According to their local environments, the distributed clouds are associated with database entries (see Sect.~\ref{sec:dirindir}). The emission from each cloud is then followed through the torus and final torus SEDs and images are obtained.}\label{fig:flowchart}
\end{figure}

In \citet{Hon06} we presented a method to time-efficiently simulate torus model images and SEDs and account for the 3D statistical nature of clumpy dust distributions. The strategy follows \citet{Nen08a}, however resolving the need for a probabilistic treatment of the dust distribution. Just as in the probabilistic approach, our model strategy is based on separating the simulations of individual cloud SEDs and images, and the final SEDs and images of the torus (see Fig.~\ref{fig:flowchart} for an illustration of our method). In a first step, the phase-angle-dependent emission for each cloud is simulated by Monte Carlo radiative transfer simulations. For that, we apply the non-iterative method delineated by \citet{Bjo01}, based on work by \citet{Luc99}. In our previous model, we used the code \textit{mcSim} \citep{Ohn06} which is able to simulate a variety of geometries. However, in order to obtain fast results for different dust compositions, we created a new Monte Carlo code which is optimized for the AGN-cloud-configuration. For each given dust composition, we determine the sublimation radius $\rsub=r(\Tsub=1\,500\,\Kel)$ from the source, and simulate several clouds at different distances (normalized for $\rsub$). To finally simulate the torus emission, dust clouds are randomly distributed around an AGN according to some physical and geometrical parameters. Each cloud is associated with a pre-simulated model cloud, after accounting for the individual cloud's direct and indirect heating balance (see Sect.~\ref{sec:dirindir}). The final torus image and SED is calculated via raytracing along the line-of-sight from each cloud to the observer. Contrary to the probabilistic clumpy torus models, this method accounts for the actual 3-dimensional distribution of clouds in the torus involving all statistical variations of randomly distributed clouds in a time-efficient way. 

\subsection{Torus parameters}\label{sec:modparams}

The distribution of clouds in the torus is initially characterized by six model parameters. The torus parameters are (1) the radial dust-cloud distribution power-law index $a$, (2) the scale height $h$ or the half-opening angle $\theta_0$, (3) the number of clouds along an equatorial line-of-sight $\No$, (4) the cloud radius at the sublimation radius $\Rclo$ in units of $\rsub$, (5) the cloud size distribution power-law index $b$, and (6) the outer torus radius $\Rout$. 
From the model parameters, several other physical properties can be derived. However, we will show that some of these parameters can be considered only marginally influential on the torus SEDs and images, so that less parameters need to be constrained by observations. In the following, we will briefly show how the model parameters are associated with the dust distribution of the torus.

The most fundamental parameters deal with the geometrical distribution of the clouds in the torus. We separate the 3-dimensional dust distribution into a radial distribution power-law, $\eta_r\propto (r/\rsub)^a$ and a vertical distribution $\eta_{z/\theta}$. For the vertical distribution, it has become common practice to use a Gaussian distribution to reproduce the smooth transition from type 1 to type 2 viewing angles as implied by observations. However, two different ways of defining this Gaussian have evolved from recent torus models. One possibility is to distribute the clouds perpendicular to the equatorial plane, i.e. in $z$-direction in a cylindrical coordinate system. The resulting distribution function $\eta_z \propto \exp(-z^2/2H^2)$ depends on the scale height $H(r)=h \cdot r$ at radial distance $r$. This implies that the torus flares with constant $H(r)/r$, as suggested by isothermal disks. Here, $h$ is the fundamental parameter describing the vertical distribution. Another way of distributing clouds can be best understood in a spherical coordinate system: instead of defining the distribution function perpendicular to the equatorial plane, it is also possible to distribute the clouds along an altitudinal path, i.e. in spherical $\theta$-direction with $\eta_\theta \propto \exp(-\theta^2/\bar{\theta_0}^2)$. Here, $\bar{\theta_0}=(\pi/2-\theta_0)$ is the half-covering angle of the torus which is the complementary of the half-opening angle $\theta_0$ (see remarks in Table~\ref{tab:func}). Most of the simulations shown in the later sections will be based on the $\theta$-distribution, but we will include results using the $z$-distribution when drawing final conclusions from the SED models.
Table~\ref{tab:func} shows the full distribution functions $\eta_r$, and $\eta_{z/\theta}$. The complete dust cloud distribution function 
\begin{equation}
\eta(r,z/\theta)=2\pi\eta_r\cdot\eta_{z/\theta}
\end{equation}
is normalized so that $\int_{1}^{\Rout}\eta_r\,\dd r = 1$ and $\eta_{z/\theta}=1$ for $z=0$ or $\theta=0$. By this definition, $\eta_r$ can be understood as the \textit{normalized} number of clouds per unit length, so that
\begin{equation}\label{eq:master}
\No\eta(r,z/\theta)=\pi\Rcl^2\,\rho_N
\end{equation}
describes the (actual) number of clouds per unit length, where $\rho_N$ denotes the cloud number density per unit volume. Note that $r$, $z$, $H$, and $\Rclo$ are given in units of $\rsub$. 

After having fixed the distribution of clouds, the total obscuration (or dust mass) of the torus has to be defined. The most convenient way to do this is by defining the mean number of clouds $\No$ that intersect the line-of-sight along a radial path in the equatorial plane. From $\No$ it is easy to calculate the total number of clouds that have to be randomly distributed in the torus,
\begin{equation}
\Ntot=\No \int\frac{\eta(r,z/\theta)}{\pi\Rcl^2(r)}\,\dd V
\end{equation}
which is essentially the integrated version of eq.~(\ref{eq:master}) \citep[see also][]{Hon06,Nen08a}. Here, $\Rcl(r) = \Rclo\,\rsub\cdot r^b$ denotes the radius of a dust clouds at radial coordinate $r$ (in units of $\rsub$), with parameters $\Rclo$ and $b$ as defined above. In Table~\ref{tab:func}, the explicit expression for $\Ntot$ is given for both vertical $z$- and $\theta$-distributions.

\begin{table}
\centering
\caption{Cloud distribution functions and relations to physical parameters of the torus}\label{tab:func}
\begin{tabular}{l l l}
\hline\hline
Physical property                & Function$^{1,2,3}$                                                     \\ \hline
radial dust distribution         & $\eta_r = \frac{1+a}{\left(\Rout^{1+a}-1\right)} \, r^a \,\cdot\,\frac{1}{\rsub}$     \\
vertical dust distribution       & $\eta_z = \exp\left(-z^2/2H^2\right)$                               \\
                                 & or $\eta_\theta = \exp\left(-\theta^2/\bar{\theta_0}^2\right)$            \\
dust cloud radius                & $\Rcl(r) = \Rclo\,r^b \,\cdot\,\rsub$                                                  \\
total number of clouds           & $\Ntot^{z} = \sqrt{8\pi} \No h \Rclo^{-2} \frac{1+a}{3+a-2b}\frac{\Rout^{3+a-2b}-1}{\Rout^{1+a}-1}$ \\
                                 & $\Ntot^{\theta} = \sqrt{2\pi} \No f_{\theta_0} \Rclo^{-2} \frac{1+a}{3+a-2b}\frac{\Rout^{3+a-2b}-1}{\Rout^{1+a}-1}$ \\
                                 & with $f_{\theta_0} = \sqrt{2}\bar{\theta_0}\exp\left(-\bar{\theta_0}^2/4\right)\,\mathrm{Erf}\frac{\pi-i\bar{\theta_0}^2}{2\bar{\theta_0}}$ \\
volume filling factor at $z=0$   & $\PhiVo = \frac{4}{3}\,\No\Rclo\,\frac{1+a}{\Rout^{1+a}-1}\cdot r^{a+b}$ \\ \hline
\end{tabular}
\begin{list}{}{\setlength{\leftmargin}{\rightmargin}}
\item --- \textit{Notes:} $^1$ $r$, $z$, $H$, and $\Rclo$ are given in units of $\rsub$; $^2$ Model parameters: (1) $a$: radial distribution power law index, (2) $h$: scale height, $h=H(r)/r=\mathrm{const.}$ or half-opening angle $\theta_0$, (3) $\No$: mean number of clouds along equatorial line-of-sight, (4) $\Rclo$: cloud radius at $\rsub$ (in units of $\rsub$), (5) $b$: cloud size distribution index, (6) $\Rout$: the outer torus radius; $^3$ We define the altitudinal angle $\theta=0$ in the torus mid-plane. Therefore we use, for convenience, the half-covering angle $\bar{\theta_0} = \pi/2-\theta_0$ instead of the half-opening angle $\theta_0$ in the equations.
\end{list}
\end{table}

Since we deal with a clumpy torus, it is most sensible to have a volume filling factor $\PhiV<1$ all over the torus. The volume filling factor $\PhiV$ can be calculated by multiplying the cloud number density per unit volume $\rho_N$ and the cloud volume $V_\mathrm{cl}=4/3\pi\Rcl^3$, so that  
\begin{equation}
\PhiV = \frac{4}{3}\No\eta(r,z/\theta)\,\Rcl\,\,\,\, .
\end{equation} 
If restricting to the equatorial plane, the expression $\PhiVo(r) = \PhiV(r,z\,\mathrm{or}\,\theta=0)$ is derived as shown in Table~\ref{tab:func}. 

We note that while consistency of our modeling approach requires $\PhiV(r,z/\theta),<1\,\forall (r,z)\,\mathrm{or}\,(r,\theta)$, the torus brightness distribution and SED is not explicitly depending on $\PhiV$ but rather on the combination of the individual parameters. Actually, approximately the same overall SEDs and images can be obtained when reducing $\PhiV$ by reducing $\Rclo$, leaving all other parameters fixed (and calculate $\Ntot$ clouds according to the new $\Rclo$). This can be illustrated by introducing a simplified version of the clumpy torus model for type 1 AGN \citep[see also][]{Kis09a}. In a face-on view of the torus, the clouds are projected onto a ring ranging from $\rsub$ to $\Rout$. First, let us assume that most of the clouds are directly heated by the AGN. Then, a ``surface filling factor'' $\sigs(r)$ can be described using cloud number density $\rho_N$ projected $z$-direction\footnote{Actually, $\sigs(r)$ can grow to larger than 1 which corresponds to the case that projection caused clouds to overlap. However, $\sigs(r)>1$ is only valid if the dust column is optically thin at that wavelength.}. If we consider $\eta_z$, the surface filling factor becomes
\begin{equation}\label{eq:surfdens}
\sigs(r) = \int_{-\infty}^{\infty} \pi \Rcl^2 \cdot \rho_N\,\dd z = \left(2\pi\right)^{3/2} \, \No h \, \eta_r \, \cdot \, \rsub
\end{equation}
with the explicit dependence $\sigs(r) \propto r^a$. In this definition, the surface filling factor can be considered as a weighting factor of how much the clouds at radius $r$ contribute to the total intensity. Here, $\No$ and $h$ are scaling constants which do not depend on the radial distribution. On the other hand, we can include a first-order approximation of obscuration effects, which accounts for a decreasing number of directly-illuminated clouds with radius, by multiplying $\sigs(r)$ with $\exp\left(-N(r,z)\right)$ as described in the next Sect.~\ref{sec:dirindir}. Finally, to obtain SEDs and brightness distributions in this simplified type-1 model, the luminosity $L_\nu^\mathrm{(tor)}$ is calculated by multiplying the surface filling factor with the source function of the clouds $S_\nu(r)$ and integrating from $\rsub$ to $\Rout$,
\begin{equation}\label{eq:master_simple}
L_\nu^\mathrm{(tor)} = 2\pi \int_{1}^{\Rout}\sigs(r)\cdot S_\nu(r)\,r\,\dd r
\end{equation}
Interestingly, $L_\nu^\mathrm{(tor)}$ does not depend on $\Rcl$ or $b$, so that it is difficult to constrain these parameters by observations. Instead, for the purpose of SED modeling, they should be selected in a way to fulfill the ``clumpy criterion'' $\PhiV<1$. 

However, since $\Rcl$ and $b$ control the total number of clouds $\Ntot$, they have influence on small scale surface brightness variations resulting in slightly different SEDs for different random arrangements of clouds \citep[see][]{Hon06} or position-angle variations, which might be interesting for interferometry (see Sect.~\ref{sec:intf}). Aside from $\Rclo$ and $b$, in Sect.~\ref{sec:rout} we will further show that $\Rout$ cannot be considered as a free parameter but has to be chosen in a sensible way depending on $a$. 

In summary, the key torus model parameters which are directly accessible from observations are $a$, $\No$, and $h$ or $\theta_0$. In addition, the cloud dust properties (dust composition and optical depth) will have some influence on the overall torus emission.

\subsection{The diffuse radiation field in the torus}\label{sec:dirindir}

As described in Sect.~\ref{sec:strategy}, our model strategy makes use of pre-calculated databases of dust clouds. The main database consists of clouds which are directly heated by the AGN. However, due to obscuration effects within the torus, some clouds will not be directly exposed to the AGN radiation since the line-of-sight to the AGN is blocked by other clouds. Such clouds are heated only by the emission from other clouds in their vicinity. These nearby clouds form a ``diffuse radiation field'' where mainly those clouds contribute to which are directly AGN-heated. In \citet{Hon06}, we used an upper-limit-approximation for the diffuse radiation which led to overestimation of the emission longward of 20$\,\micron$. Here we describe a way to recover the diffuse radiation field statistically.

For given distance from the AGN and optical depth of each cloud, the parameters which determine the temperature and emission of a cloud are (1) the fractional cloud area, $\fDH$, which is directly heated by the AGN, and (2) the fraction of directly-heated clouds in the cloud's vicinity, $\fIH(r)$. $\fDH$ determines the strength of direct heating. For $\fDH=1$, the directly-heated re-emission of the cloud is equal to the emission of the pre-calculated database cloud at the same distance $r_\mathrm{DB} = r_\mathrm{cl}$ from the AGN. If $\fDH<1$, less direct energy is received and the cloud emission corresponds to a database cloud at distance $r_\mathrm{DB} = r_\mathrm{cl}/\sqrt{\fDH}$. The parameter $\fIH(r)$ determines the energy that is contained in the diffuse radiation field. It can be approximated by the probability $\Pfoc$ that a given cloud at $r$ and $z$ (or $\theta$) is directly illuminated. Note that $\Pfoc$ can be considered a global, probabilistic version of $\fDH$, which is a local and individual property of each cloud. According to \citet{Nat84}, $\Pfoc$ can be derived from Poisson statistics as $\Pfoc = \exp\left(-N(r,z/\theta)\right)$, where $N(r,z/\theta)=\No\int_0^s \eta(r,z/\theta) \dd s$ is the mean number of clouds along the path $s$ from the center to $(r,z/\theta)$. In case $\fIH\equiv\Pfoc=1$ (and $\fDH=0$), the cloud's contribution from indirect heating corresponds to the indirectly-heated database cloud emission at the same distance, $r_\mathrm{DB} = r_\mathrm{cl}$, otherwise $r_\mathrm{DB} = r_\mathrm{cl}/\sqrt{\fIH}$.

\begin{figure}
\centering
\includegraphics[width=.5\textwidth]{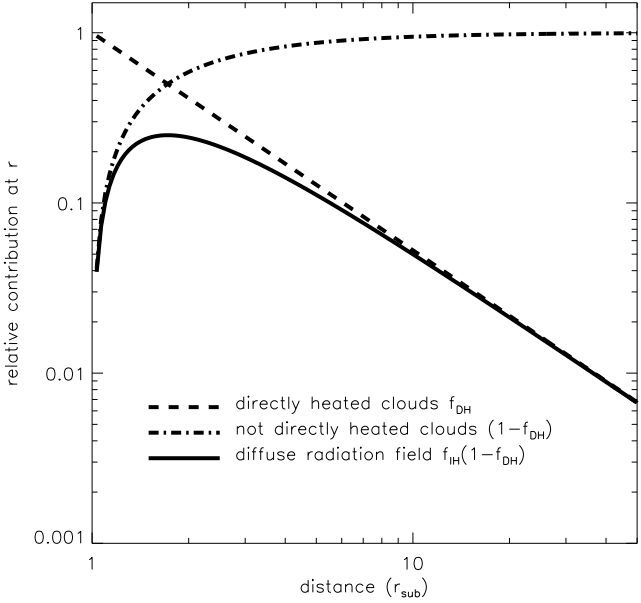}
\caption{Illustration of the contribution of directly- and indirectly-heated clouds. The dashed line represents the relative contribution of directly-heated clouds to the torus emission at given distance $r$ from the AGN in the mid-plane, $\fDH$. The dash-dotted line shows the relative contribution $\left(1-\fDH\right)$ of indirectly-heated clouds at $r$ if $\fIH=1$. The solid line presents the actual contribution of indirectly-heated clouds due to the decreasing strength of the diffuse radiation field, $\fIH(1-\fDH)$.}\label{fig:soccont}
\end{figure}

An interesting aspect is the actual contribution of indirectly-heated clouds to the overall torus emission. On one hand, with increasing $r$ the fraction of clouds which are indirectly heated increases due to the increase in obscuration. On the other hand, the diffuse radiation field becomes weaker due to the absence of directly-heated clouds. In Fig.~\ref{fig:soccont}, we illustrate the relative contribution of directly- and indirectly-heated clouds to the torus emission at a given distance $r$ in the torus mid-plane (i.e. $z=0$ or $\theta=0$). Here, we fix $\No=5$, $a=-1.0$, and $\Rout=50$\footnote{For $a=-1.0$, the normalized radial distribution function becomes $\eta_r = (r\ln\Rout)^{-1}$.}. The relative contribution, $\fDH$, of directly-heated clouds at distance $r$ from the AGN is shown as a dashed line in Fig.~\ref{fig:soccont}. As a dashed-dotted line, we show the relative contribution of indirectly-heated clouds if we assume that the vicinity of the indirectly-heated cloud is fully filled-up with directly-heated clouds to form the diffuse radiation field, i.e $(1-\fDH)$. This was assumed as the contribution of indirectly-heated clouds in \citet{Hon06}, leading to the overestimation of the SEDs at wavelengths $>20\,\micron$ \citep[see Appendix A.2 in][]{Hon06}. In reality, the number of directly-heated clouds around an indirectly-heated cloud decreases with $r$, as outlined in the previous paragraph. The solid line illustrates this effect. It shows the contribution of indirectly-heated clouds at distance $r$, considering the decrease of directly-heated clouds in the vicinity, $\fIH(1-\fDH)$. As can be expected from energy conservation, this line is lower than the directly-heated cloud contribution and approaches it asymptotically with increasing $r$. Given their general higher temperatures, this illustrates that directly-heated clouds dominate the torus emission over indirectly-heated clouds -- at least in more or less face-on geometries where line-of-sight obscuration effects play a minor role. 

Finally, we want to justify why a single iteration of the diffuse radiation field is sufficient. Since the directly heated clouds are also indirectly heated, one would first have to simulate directly-heated clouds. From these clouds, the diffuse radiation field has to be calculated and be included in the simulations of directly-heated clouds. Then, a new diffuse radiation field has to be determined and the process be repeated. After several iteration steps, the final diffuse radiation field is obtained. The Monte Carlo method we used is quite flexible in implementing such a scheme without the need to reconsider direct illumination or previous steps \citep[][p.283]{Kru08}: After the $i$-th simulation, a ``corrected'' diffuse radiation field has to be added which corresponds to the change with respect to the previous $(i-1)$-th iteration step, $\dd S^{(i)}_{\nu;\mathrm{DRF}} = S^{(i)}_{\nu;\mathrm{DRF}}-S^{(i-1)}_{\nu;\mathrm{DRF}}$. However, it has been noted that the diffuse radiation field converges quite fast \citep[e.g.][]{Hon06,Nen08a}, so that one can consider that the source function of the diffuse radiation field $S_{\nu;\mathrm{DRF}} \approx S^{(0)}_{\nu;\mathrm{DRF}} = \int_0^\pi S_{\nu;\mathrm{DH}}(\phi)\sin\phi\,\dd\phi$, where $S_{\nu;\mathrm{DH}}$ is the source function of (AGN-only) directly-heated clouds, without the need of any further iteration. This makes simulations of cloud databases very time efficient.

\section{Emission from individual dust clouds}\label{sec:emcld}

In this section we will discuss dust cloud modeling results from step I of the torus simulation (see Fig.~\ref{fig:flowchart}). First, we discuss physical properties of the dust clouds in the torus. Next, we detail the Monte Carlo simulation of individual dust clouds and explain underlying assumptions. Finally we present cloud SEDs for both directly- and indirectly-heated dust clouds for different dust compositions.

\subsection{Properties of the dust clouds}\label{sec:cloud_prop}

\subsubsection{Observational and theoretical constraints on physical properties}

One fundamental property of the dust clouds in radiative transfer simulations is their optical thickness. If they are optically thin, i.e. $\tau_V\ll1$, then they could be considered to have uniform temperature and the thermally re-emitted flux can be approximated by $\Fnu = \pi \tau_\nu B_\nu(T)$, where $B_\nu(T)$ is the Planck function at temperature $T$ and $\tau_\nu$ is the frequency-dependent optical depth. However, from observations of type 2 AGN, we know that both the dust and hydrogen columns are usually quite large, with the hydrogen column even Compton-thick. Associated torus dust temperatures are rather cool, peaking at $T\sim200-300\,\Kel$ \citep[e.g.][]{Jaf04,Tri07}. Moreover, X-ray column density variability as seen in many AGN point to rather small clouds and column densities of the order of $10^{23}\,\mathrm{cm^{-2}}$ per cloud \citep[e.g.][]{Ris07}. To combine silicate absorption features as seen in type 2 AGN, Compton-thick obscuration, and small clouds, we infer that there are on average about 5$-$10 optically-thick clouds ($\tauV\gg1$) along a line-of-sight to the center in a type 2 AGN and that the volume filling factor of the torus $\PhiV<1$, i.e. the torus is clumpy. These cloud properties are, at least qualitatively, in agreement with models of self-gravitating dust and gas clouds in the shear of the gravitational potential of a super-massive black hole \citep{Bec04,Hon07b}.

\subsubsection{Dust composition}\label{sec:dustcloud}

\begin{table}[t]
\caption{Main physical properties of the dust compositions used in our study.}\label{tab:dust}
\begin{tabular}{l | c c | c c | c}
\hline\hline
Name identifier   & \multicolumn{2}{c |}{fractional composition} & $\amin$   & $\amax$     & $\rsub$$^a$ \\
                  & silicates         & graphite              & ($\micron$) & ($\micron$) & (pc)        \\ \hline
standard ISM      & 0.53               & 0.47                  & 0.025       & 0.25        & 0.9         \\
ISM large grains  & 0.53               & 0.47                  & 0.1         & 1.0         & 0.5        \\
Gr-dominated      & 0.30               & 0.70                  & 0.05        & 0.25        & 0.7        \\ \hline
\end{tabular}
\begin{list}{}{\setlength{\leftmargin}{\rightmargin}}
\item --- {\it Notes:} $^a$ Sublimation radius $\rsub$ at a dust temperature $\Tsub = 1\,500\,\Kel$ for $\Lbol=10^{46}\,\ergs$. Sublimation radii for any other luminosity can be calculated by $r = \rsub \times (\Lbol/10^{46}\,\ergs)^{1/2}$. $\Lbol$ is defined as $\Lbol=\int\Lnu\dd\nu$ of the accretion disk. In our definition of the accretion disk spectrum, the reverberation radius from \citet{Kis07} for $\Lbol=10^{46}\,\ergs$ would be $\rrev=0.38\,\pc$, with approximately 25\% scatter.
\end{list}
\end{table}

Previous torus studies used dust properties in line with standard grain sizes and chemical composition \citep[e.g.][]{Sch05,Hon06,Nen08b}. In general, MRN distributions \citep[referring to Mathis, Rumpl \& Nordsieck;][]{Mat77} with a grain size power law distribution $\propto a^{-3.5}$ ($a\ldots$grain size) was assumed with lower and upper limits $\amin\sim0.005-0.01\,\micron$ and $\amax\sim0.25\,\micron$, respectively. The chemical compositions invoked 47\% graphite and 53\% silicates, either based on \citet{Dra03} or \citet[][for silicate dust]{Oss92} optical properties. Such dust compositions are considered as ``standard ISM''. Recently it was suggested that the dust composition around AGN might be deviating from standard ISM. While \citet{Sug06} confirmed the $L^{1/2}$-dependence of the sublimation radius, the measured $K$-band reverberation radii of type 1 AGN are a factor of 3 smaller than what is expected from standard ISM dust. \citet{Kis07} suggest that this discrepancy might be caused by domination of larger grains, at least in the inner part of the tori. 

We aim for testing the impact of different dust compositions on the cloud SEDs. We synthesize standard ISM mixes invoking 47\% graphite (using optical constants from \citet{Dra03}) and 53\% silicates using both \citet{Dra03} and \citet{Oss92} optical constants. In order to explore recent suggestions, we also create a dust mix with standard ISM grain size distribution but only containing grains between 0.1\,$\micron$ and 1\,$\micron$ in size (``ISM large grains''). A last composition explores SEDs of dust which is dominated by intermediate to larger graphite grains (70\% graphite, 30\% silicates; ``Gr-dominated''). The compositions and their respective properties are summarized in Table~\ref{tab:dust}.

\subsection{Monte Carlo simulations of dust clouds}\label{sec:cloudMC}

Since the exact dust distribution within a cloud is not known, we use a uniform density throughout the cloud. While this might be oversimplified, we note that independent of the actual density distribution of an optically thick cloud, the bulk of the cloud re-emission originates from the optically-thin surface layer of the hot, directly-illuminated face of the cloud. Moreover, the gradient of the temperature at distance $r$ from the cloud surface depends mostly on the total optical thickness at $r$ and is supposedly very similar in all density distribution laws once the same optical depth is reached. So, the temperature of the non-illuminated side of the cloud is more or less independent of the dust density law within the cloud for the same optical depth through the diameter of the cloud (=``total optical depth'' $\tauV$). 

The cloud is subdivided into grid cells, where the dust density within each cell is optically thin. This guarantees that the steep temperature gradient in the surface layer on the hot face of the cloud is properly sampled.

For our cloud simulation, we use a spherical cloud geometry. In principle any kind of shape could have been chosen. However, the main difference between a sphere and, say, an ellipse is the different ratio of radial-to-perpendicular optical depth. As mentioned above, if the cloud has a total $\tau_V\gg1$, then the dominant emission is coming from the optically-thin layer on the hot side of the cloud. Since the temperature in the layer is mostly independent of the density distribution in the rest of the cloud, the overall temperature of the hot cloud face is similar. For the cold face of the cloud, the main criterion is again the total optical depth within the cloud. For comparable total optical depths of two clouds with arbitrary shape\footnote{The actual total $\tau_V$'s that need to be compared depend on the exact shapes/elongations. Thus, one might see the total $\tauV$ of the cloud in our model as some mean value representing all kind of shapes.}, the cold-side temperatures are the same and, thus, the emerging emission. In summary, unless one insists on very peculiar cloud surfaces, the exact shape does not matter too much and a spherical cloud presumably catches the essence of the dust clouds in AGN tori.

For time-efficient simulations, we optimize the Monte Carlo code for the cloud-AGN-configuration. For that, we first determine the sublimation radius $\rsub=r(\Tsub)$ for a given dust chemistry and grain size composition, assuming a dust sublimation temperature $\Tsub = 1\,500\,\Kel$. Since $\rsub$ scales with the AGN luminosity $L$ as $\rsub \propto L^{1/2}$, the incident AGN flux on a dust cloud at $\rsub$ is independent of the actual luminosity, $F_\mathrm{AGN}(\rsub) \propto L/\rsub^2 = \mathrm{constant}$. Thus, a single cloud database can be used for all AGN luminosities. Since the dust clouds are believed to be small, with a radius-to-distance-ratio $\Rcl/r\la1$, we assume that the incident AGN radiation (see Sect.~\ref{sec:agn_radfield}) enters the model sphere on parallel rays. Moreover, the rotational symmetry of our clouds allows to reduce the model space for Monte Carlo simulations to a planar 3-dimensional grid with only 1 grid cell in $z$-direction. These two modification allow for a maximum number of photon packages to be simulated in a short time without suffering from strong Monte Carlo noise in the temperature of the individual grid cells. To be even less dependent on Monte Carlo noise, the final cloud SEDs and images are calculated via raytracing of the emission from each cell into the observer's direction (i.e. for different phase angles of the cloud): First, the temperature distribution is reconstructed from the temperature grid; then, the source function (thermal re-emission and scattering) for each cell is calculated and the path through the cloud is followed out of the cloud\footnote{Note that this treatment limits the handling of scattering to single events. For our optically-thick clouds we tested this method against rigorous treatment of scattering with the \textit{mcsim} code and found that any noticeable impact of the single-event-assumption is limited to wavelengths $\la1\,\micron$ and is of the order of $2\%$.}. In this way the angle-dependent source function of the cloud $S_\nu(\phi)$ is time-efficiently recovered. The described simulation method is the same for both directly- and indirectly-heated clouds. The only difference occurs in handling the incident radiation: in the case of directly-heated clouds, the AGN spectrum is used while for indirectly-heated cloudes the diffuse radiation field is the heating source.

\begin{figure*}
\centering
\includegraphics[width=1.0\textwidth]{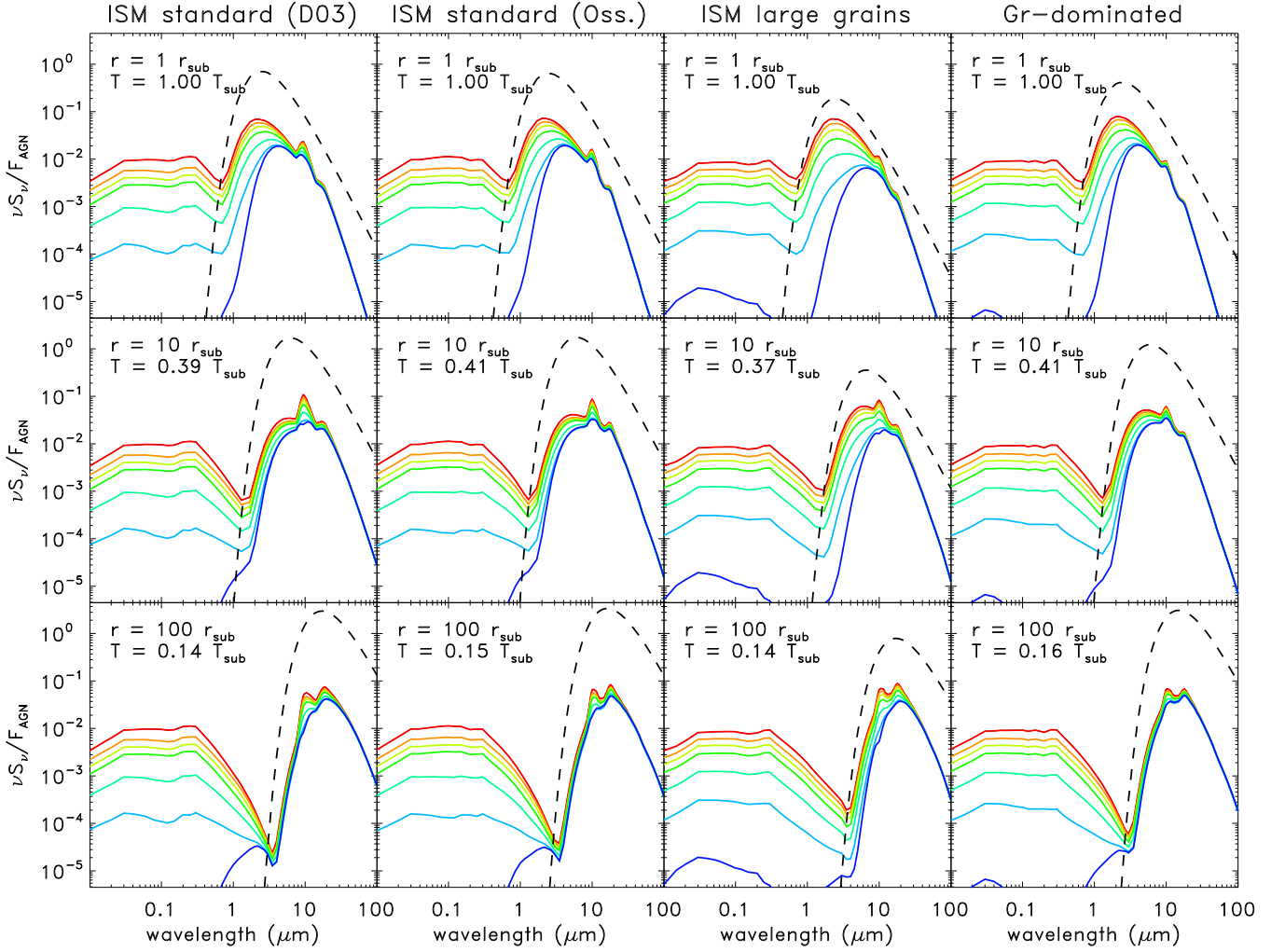}
\caption{Model source functions $S_\nu$ of directly AGN-heated clouds for several different dust configurations (from left to right column: ISM standard configuration with Draine (2003) silicates, Ossenkopf et al. (1994) silicates, ISM large grains, and Graphite-dominated dust). Each row shows the emission of a cloud with $\tauV=50$ at the given AGN distance $r=\rsub$ (top), $10\,\rsub$ (middle) and $100\,\rsub$ (bottom), scaled to the total incoming flux of the AGN, $F_\mathrm{AGN} = \Lbol/(4\pi r^2)$. The corresponding hot-side temperature in units of $\Tsub=1\,500\,\Kel$ are also given. The various lines represent a specific phase angle, $\phi$, of the cloud in steps of $30\deg$ (red: $\phi=0\deg$, i.e. full hot side; purple: $\phi=180\deg$, i.e. full cold side). For comparison, we overplotted the corresponding black-body emission $B_\nu(T)$ with the same temperature as the hot side of the cloud at the same distance from the AGN (dashed line).}\label{fig:dustclouds}
\end{figure*} 

In summary, our simulations of dust clouds proceeds as follows (see also \citet{Hon06,Nen08a}): (1) Monte Carlo simulations of directly AGN-heated clouds; (2) Approximation of the diffuse radiation field (i.e. heating of clouds not directly exposed to the AGN) by averaging the local emission of directly-heated clouds (see also Sect.~\ref{sec:dirindir} for a description of the assessment of the local circumstances within the torus); (3) Monte Carlo simulation of clouds heated by the averaged diffuse radiation field. This results in two cloud databases which are used for the torus simulations: one for directly-illuminated clouds and one for indirectly-heated clouds.

\subsection{Resulting cloud SEDs}

In the following we present model source functions of directly- and indirectly-heated clouds. Simulations have been carried out for the standard-ISM, ISM large-grains, and Gr-dominated dust configurations as introduced in Sect.~\ref{sec:dustcloud}.

\subsubsection{Directly-heated clouds}

\paragraph{Standard ISM}

The starting point for our dust composition study will be the ``standard ISM''. Physical dust properties are listed in Tab.~\ref{tab:dust}. We show two versions of the standard ISM configuration: one uses \citet{Dra03} silicates, the other one consists of \citet{Oss92} silicates (see Sect.~\ref{sec:dustcloud}). The most important difference between both silicate types concerns the central wavelength and width of the silicate feature at around 10\,$\micron$. For a given size distribution, the Draine silicate feature is much wider than the Ossenkopf feature. In Draine silicates, the central silicate feature wavelength is at $\sim9.5\,\micron$ while the Ossenkopf silicate feature is centered at $10.0\,\micron$. The $10.0\,\micron$ central wavelength seems to be more consistent with peak silicate emission features of type 1 AGN as observed with \Spitzer, while absorption features in type 2s usually have their center at $9.7\,\micron$ \citep[e.g.][]{Shi06}. We note, however, that radiative transfer effects can affect the exact shape of the feature \citep[but see discussions in][which disagree on the details of this effect]{Nik09,Hon10,Lan10}.

\begin{figure*}
\centering
\includegraphics[width=0.85\textwidth]{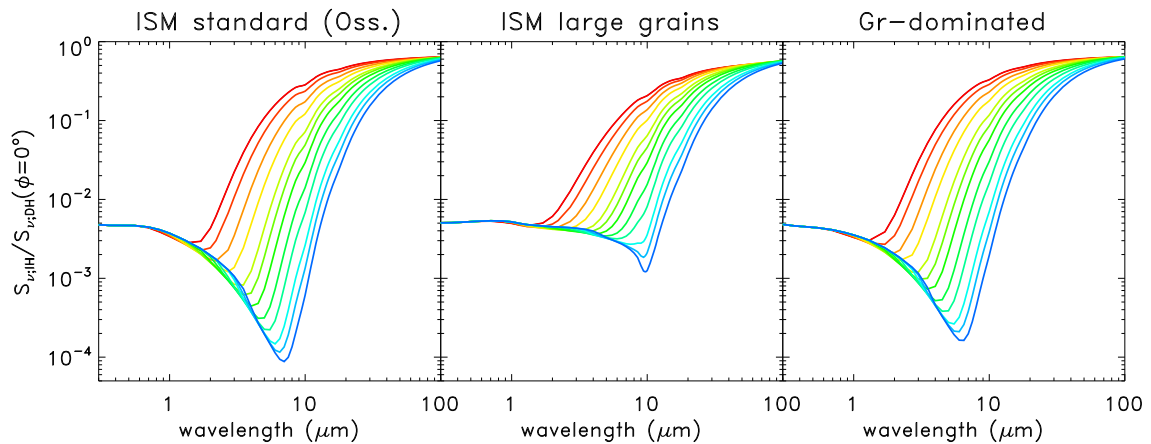}
\caption{Wavelength-dependent emission ratio between the source function of indirectly-heated clouds $S_{\nu\mathrm{;IH}}$ and the hot face of directly-heated clouds $S_{\nu\mathrm{;DH}}(\phi=0^\circ)$. The various lines represent different distances $r$ from the AGN with $r$ = 1, 1.5, 2.5, 4, 7, 10, 15, 25, 40, 70, 100 and 150\,$\rsub$ from top-red to bottom-blue. The left panel shows simulation results for ``ISM standard'' dust with Ossenkopf et al. silicates, the middle panel presents ``ISM large grains'', and the right panels shows ``Gr-dominated'' dust.}\label{fig:socclouds}
\end{figure*}

Simulation results for directly-illuminated clouds with standard ISM dust configurations are shown in the two left columns of Fig.~\ref{fig:dustclouds}. We present clouds with a total optical depth $\tauV=50$ at three different distances $r=\rsub$, $10\times\rsub$, and $100\times\rsub$ from the AGN. Along with the distances, we also provide the maximum temperature of the hot side of the cloud in units of the sublimation temperature $\Tsub = 1\,500\,\Kel$. The overplotted dashed lines show the black-body source function $B_\nu(T)$ with the same temperature as the hot side of the clouds at the same distance from the AGN. For both types of silicates, the temperature gradient on the directly-illuminated face is similar, with some trend of slightly higher temperatures for the Ossenkopf dust. The $r=10\times\rsub$ clouds serve best for illustration since the near-IR color is slightly redder for Draine silicates. The different temperature gradients are a result of slightly different absorption efficiencies -- or more precisely: Planck mean opacities which, at the end, determine the temperature at a given radius in the radiative transfer equations. The cloud emission is lower than the corresponding black-body emission by more than a magnitude indicating that ISM standard mixture clouds are not well represented by any black-body approximation. This is true for all dust compositions discussed here. It is worth noting that for the standard ISM configuration, the presented clouds with $\tauV=50$ are mostly optically thin in the infrared. This leads to only small differences in SED shapes for different phase angles $\phi$\footnote{The phase angle $\phi$ is defined as the AGN-cloud-observer angle. $\phi=0\deg$ denotes the full visibility of the directly-illuminated face of the cloud (``full moon''), $\phi=180\deg$ describes the situation when only the cold side of the cloud is seen (``new moon'').} so that the hot-side emission shines all through the cloud and significantly contributes to the cold-side emission, e.g. as seen by the 9.7/10\,$\micron$ silicate emission feature in the $\phi=180\deg$ SEDs.

\paragraph{ISM large grains}\label{sec:largegrains}

While near-IR reverberation mapping results of type 1 AGN confirmed the $L^{1/2}$-dependence of $\rsub$, the observed $K$-band reverberation mapping radii $\rrev$ are smaller by approximately a factor of 2$-$3 than what has been inferred from ISM dust \citep{Sug06,Kis07}. Suggestions to solve this problem involve larger sizes of the dust grains than used in standard ISM distributions -- at least in the innermost part of the torus which dominates the IR emission of type 1 AGN. To test this suggestion, we simulated dust clouds for a standard ISM composition with Ossenkopf silicates and limited the grain sizes to $0.1-1\,\micron$. This ``ISM large grain'' composition has a sublimation radius $\rsub = 0.5\,\pc\times(\Lbol/10^{46}\,\ergs)^{1/2}$ which is closer to the observed $\rrev$ than the standard ISM dust (see Tab.~\ref{tab:dust}). 

The simulated dust cloud SEDs for the ISM large grain composition are shown in the third column of Fig.~\ref{fig:dustclouds}. The hot-side temperature gradient is comparable to clouds with ISM standard dust. The most notable difference concerns the change of the SED shape from $\phi=0\deg$ to $180\deg$. While the previously-explored clouds show quite similar SEDs for all phase angles, the ISM large grain clouds become very red at near- and mid-IR wavebands for increasing $\phi$. Moreover, the silicate feature at $10\,\micron$ turns from emission into absorption. The reason for this difference is the optical depth in the infrared: For both silicate and graphite, large grains have a much flatter extinction efficiency in the infrared. Thus, the cloud-internal extinction in the IR increases. While $\tau_{2.2\,\micron}/\tauV \sim 0.05-0.06$ for the ISM standard configuration (depending on the silicate species), the ISM large grains have $\tau_{2.2\,\micron}/\tauV \sim 0.9$. In the silicate feature at $9.7/10\,\micron$, the ratios are $\tau_\mathrm{Si}/\tauV\sim0.04$ and $0.1$, respectively. As a consequence, a cloud with $\tauV=50$ is also optically thick in the near- and mid-infrared. Thus, when seeing the cloud under large $\phi$, the emission from the hot side of the cloud is blocked and the emerging SED is dominated by cool dust emission. This will also have an effect for the simulated torus SED since IR radiation can be efficiently absorbed which presumably leads to redder IR colors (see Sect.~\ref{sec:raddistr}). 

We note that the difference between black-body and hot-side cloud emission is smaller than for the other dust composition, as seen in Fig.~\ref{fig:dustclouds}. This implies that large dust grains are better represented by a black-body approximation than smaller grains, although the difference is still almost an order of magnitude.

\paragraph{Graphite-dominated dust}\label{sec:grdom}

In general, silicate dust grains have lower sublimation temperatures than graphite grains \citep[e.g.][for simulations with AGN radiation]{Sch05}. This might result in a dearth of silicates in the torus, at least in the inner part. Moreover, since AGN are rather strong X-ray emission sources, it is doubtful that smallest dust grains can easily survive without being photo-destructed. For that, we modified the ISM composition by using only grains $>0.05\,\micron$ and changing the chemical mixture to 70\% graphite grains and 30\% Ossenkopf silicates.

In the right-most column of Fig.~\ref{fig:dustclouds}, we show simulated cloud SEDs for this ``Gr-dominated'' dust. The overall SED shape resembles those of the standard ISM configurations. The dust is optically thin in the IR with $\tau_{2.2\,\micron}/\tauV=0.06$ and $\tau_\mathrm{Si}/\tauV = 0.02$. However, two important differences can be seen. First, the silicate emission feature at $10\,\micron$ is significantly reduced due to the reduction of silicate dust. This might be interesting given the relative weakness of silicate absorption and emission features observed in Seyfert galaxies (but see Sect.~\ref{sec:raddistr} for actual simulations). Second, the temperature gradient outward from $\rsub$ is slightly flatter. As can be seen in the middle panel, the near-IR emission in the Gr-dominated dust clouds is bluer than for the standard ISM configurations and the ISM large grains. Therefore, it can be expected that for a given set of torus parameters (see Sect.~\ref{sec:modparams}), the radial emission size in the near- and mid-IR is slightly different than for the other dust compositions.

\begin{figure*}
\centering
\includegraphics[width=1.0\textwidth]{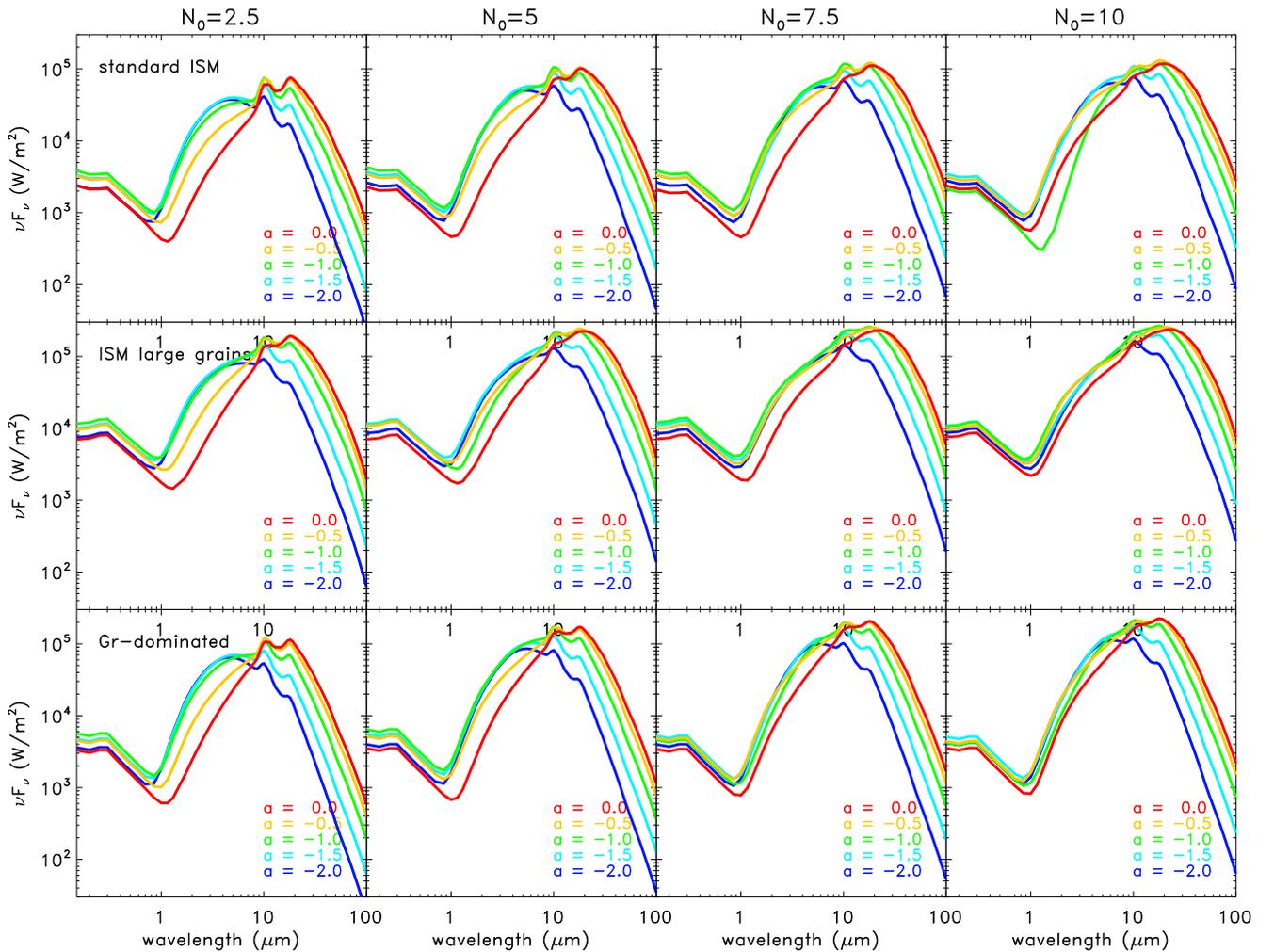}
\caption{Model SEDs of a face-on type-1 AGN (inclination angle $i=0\deg$). The top row shows models with the standard ISM composition, the middle row is for large grains, and the bottom row represents GR-dominated dust. From left to right, the columns show an increasing mean number of clouds in the equatorial line-of-sight, $\No=2.5$, $5$, $7.5$, and $10$, respectively. In each panel, we plot SEDs of one random cloud distribution for radial power law indices $a=0.5$ (red), $-0.5$ (orange), $-1.0$ (green), $-1.5$ (light blue), and $-2.0$ (dark blue). Fixed parameters are $\Rout=150$, $\theta_0=45^\circ$, $\Rclo=0.035$, $b=1$, and $\taucl=50$.}\label{fig:tor_rad_N}
\end{figure*}

\subsubsection{Indirectly-heated clouds}\label{sec:indheat}

In Fig.~\ref{fig:socclouds}, we illustrate the contribution of indirectly-heated clouds. The left panel shows the ratio of indirectly-heated cloud emission to directly heated emission as the ratio of the source functions $S_{\nu;\mathrm{IH}}/S_{\nu;\mathrm{DH}}(\phi=0\deg)$ for the ISM standard dust with Ossenkopf silicates at different distances $r$ from the AGN. In the middle panel, we present results for the ISM large grain configuration, and in the right panel Gr-dominated dust was used. Except for the clouds closest to the AGN ($r\la5\times\rsub$), the emission contribution of indirectly-heated clouds in the near- and mid-infrared below $10\,\micron$ is $\ll0.1$ for the ISM standard and Gr-dominated dust. This means that indirectly-heated clouds are much cooler than the corresponding directly-heated cloud at the same $r$. This is slightly different for the ISM large grain configuration: Since these clouds are also optically thick in the infrared, they absorb the incident diffuse radiation field -- which is dominated by IR photons -- much more efficiently than the other two dust configurations. This results in better indirect heating so that the temperatures become higher. 

\begin{figure*}
\centering
\includegraphics[width=1.0\textwidth]{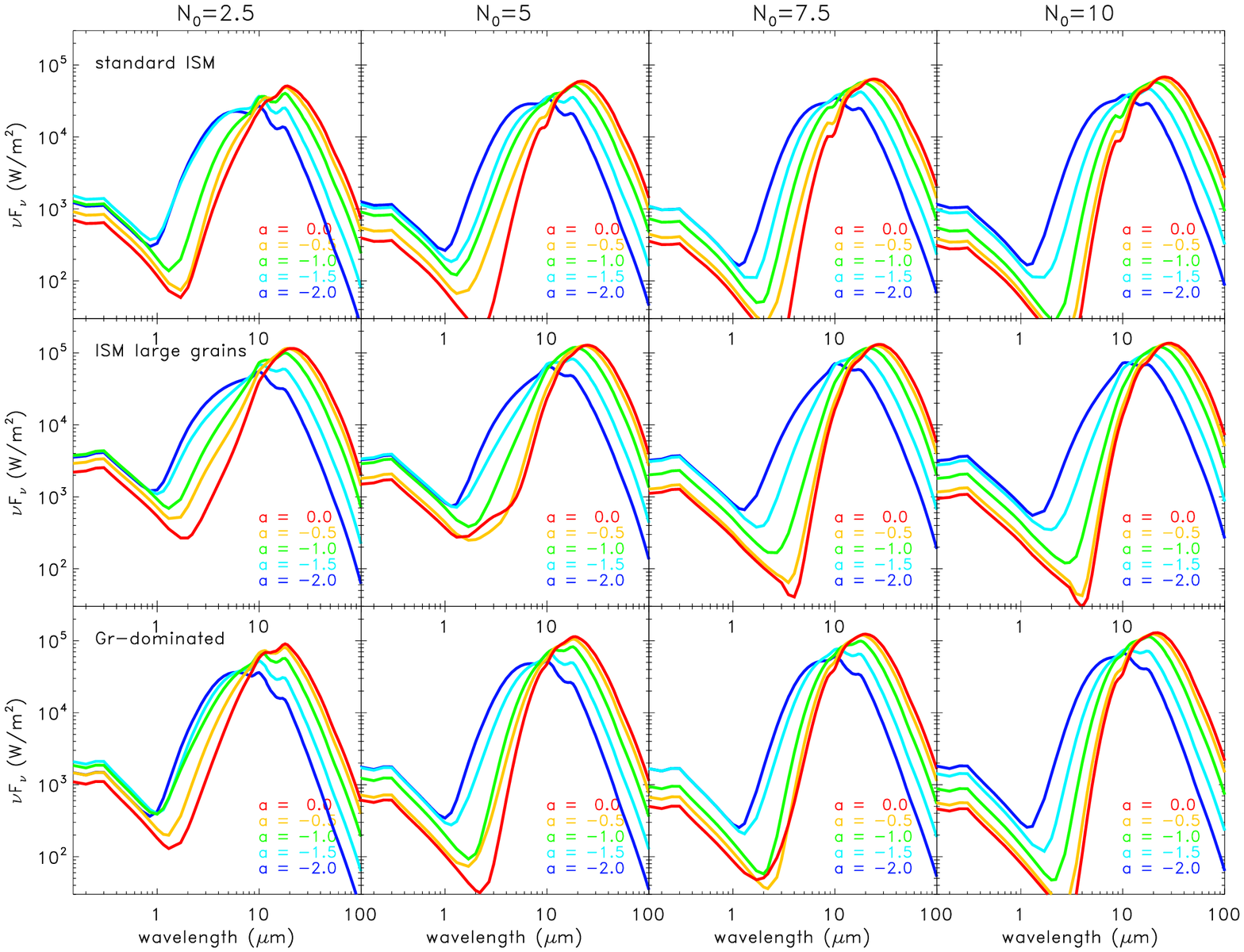}
\caption{Model SEDs of an edge-on type-2 AGN (inclination angle $i=90\deg$). The top row shows model models with the standard ISM composition, the middle row is for large grains, and the bottom row represents GR-dominated dust. From left to right, the columns show an increasing mean number of clouds in the equatorial line-of-sight, $\No=2.5$, $5$, $7.5$, and $10$, respectively. In each panel, we plot SEDs of one random cloud distribution for radial power law indices $a=0.5$ (red), $-0.5$ (orange), $-1.0$ (green), $-1.5$ (light blue), and $-2.0$ (dark blue). Fixed parameters are $\Rout=150$, $\theta_0=45^\circ$, $\Rclo=0.035$, $b=1$, and $\taucl=50$.}\label{fig:tor_rad2_N}
\end{figure*}

\section{Torus model results and discussion}\label{sec:resdis}

In this section we will present results from the full torus simulations as described in Sect.~\ref{sec:strategy} and illustrated in Fig.~\ref{fig:flowchart}. There have already been a number of SED parameter studies on clumpy torus models in literature \citep{Dul05,Hon06,Sch08,Nen08b}, so that we want to focus on the dust distribution and how it connects to observations. In particular, we will show that certain SED and interferometric properties are almost exclusively depending on the radial distribution of the dust clouds in the torus. The $\theta$-distribution will be used to discuss parameter dependencies, but the $z$-distribution will be included in the conclusions drawn from the SED studies in Sect.~\ref{sec:sed_sum}. Both distributions yield very similar results.

\subsection{Torus Spectral Energy Distributions}

All torus SEDs will be scaled independent of AGN luminosity and spatial scaling. The torus model SEDs $\nu F_\nu$ shown in this section represent the emission at a distance corresponding to the sublimation radius $\rsub$ of the respective dust composition. Observed fluxes $\nu f_\nu$ can be easily compared to model fluxes $\nu F_\nu$ via the relation
\begin{equation}
\nu F_\nu = \nu f_\nu \cdot \left(\frac{D_L}{\rsub}\right)^2
\end{equation}
where $D_L$ is the luminosity distance to the AGN and $\rsub$ is the sublimation (or reverberation) radius for the modeled dust composition which can be taken from Table~\ref{tab:dust}. By that, a comparison of observed and model SEDs allows for constraining the sublimation radius from which, if needed, the bolometric luminosity can be inferred (see Table~\ref{tab:dust}). 

\subsubsection{The radial distribution of dust clouds}\label{sec:raddistr}

In the parametrization of the torus the radial distribution of the clouds takes a key role in determining the emission, as illustrated in eqs. (\ref{eq:master}), (\ref{eq:surfdens}), and (\ref{eq:master_simple}). This is reflected by the differences in simulated SEDs when varying the power-law index $a$ in the radial dust-cloud-density distribution $\eta_r \propto r^a$. In Figs.~\ref{fig:tor_rad_N} \& \ref{fig:tor_rad2_N} we present comparison of model SEDs for different radial power-law indices $a=0.0$, $-0.5$, $-1.0$, $-1.5$, and $-2.0$, indicated by colored lines in the plot. Fig.~\ref{fig:tor_rad_N} represents a face-on line-of-sight onto the torus as in type 1 AGN ($i=0^\circ$ while Fig.~\ref{fig:tor_rad_N} shows SEDs for a type 2 case (i.e. face-on; $i=90^\circ$). Each row represents a different dust composition as discussed in Sect.~\ref{sec:cloud_prop} (top row: Ossenkopf et al. standard ISM; middle row: ISM large grains; bottom row: Gr-dominated dust). Each column shows variations of the mean number of clouds in the equatorial line-of-sight $\No$ (see Sect.~\ref{sec:modparams}), increasing from left to right ($\No=2.5$, $5$, $7.5$, and $10$). Other fixed parameters are $\Rout=150$, $\theta_0=45^\circ$, $\Rclo=0.035$, $b=1$, and $\taucl=50$.

All models have in common that the total SED becomes redder when the power-law becomes flatter, simply because cooler dust at larger distances is involved with the flat power-law distributions. The standard ISM and Gr-dominated dust distributions show similar overall SEDs and similar trends when changing $\No$ or $a$. For small $\No$, the silicate emission features are strongly pronounced, and there is quite some difference in the continuum shape for different $a$. While the flat distributions ($a=0.0$ to $-1.0$) peak in the continuum at long mid-IR wavelengths, the steeper distributions ($a=-1.5$ and$-2.0$) have their continuum emission peak in the near-IR, which is caused by the fact that in the latter case most of the dust is confined to small distances from the AGN, i.e. the average dust temperature is rather high. This is the same trend for type 1 and type 2 line-of-sights, although type 2 SEDs appear slightly redder than their type 1 counterparts.

A special case in this study are SEDs for the ISM large grain dust. While they follow the general trend of redder SEDs for shallower dust distributions, the peak continuum emission is located at wavelengths $>10\,\micron$ in $\nuFnu$ for \textit{all} selections of $a$ and $\No$ in the type 2 case, and their continuum is generally redder in type 1 orientations. 
The reason for this behavior is the dust-specific $\tau$ in the infrared (see Sect.~\ref{sec:largegrains}). For the clouds with total $\tauV=50$, the large grains are optically thick in the near- and mid-IR ($\tau_\mathrm{IR} > 1$), while the other dust compositions are optically thin in the mid-IR. This can also lead to stronger statistical effects, depending on the actual distribution of clouds \citep[see][]{Hon06}. Considering overall similarities in the SEDs and in the silicate features of most type 1 cases, it seems difficult, however, to constrain the chemistry by SED observations.

While the overall SED shows considerable dependence on $a$, it does not change significantly when varying $\No$ for both type 1 and type 2 line-of-sights. There is only a small change in color from $\No=2.5$ to 10 which can be noted as a smaller dispersion between the extreme $a$-cases by comparing the left-most with the right-most panels in Figs.~\ref{fig:tor_rad_N} \& \ref{fig:tor_rad2_N}. In fact, the red continuum in flat distributions becomes bluer and the blue, steep dust distribution become redder. This is caused by obscuration effects within the torus: if $\No$ increases, clouds at larger distances contribute less to the overall emission since there is less chance that they are directly exposed to AGN emission. That effect is strongest for the flat distributions. In total, however, torus-internal obscuration effects on the SED shape, originating from the selection of $\No$, are minor. This can be taken as a justification of the simplified clumpy torus model for type 1 AGN as introduced by \citet{Kis08} and quantified in eq. (\ref{eq:master_simple}) which captures the essence of the (overall) brightness distribution in pole-on geometries. However, the selection of $\No$ seems to have an impact on the strength of silicate feature, which will be discussed in the next section.

\subsubsection{Average number of clouds along an equatorial line-of-sight}\label{sec:n0}

\begin{figure*}
\centering
\includegraphics[width=1.0\textwidth]{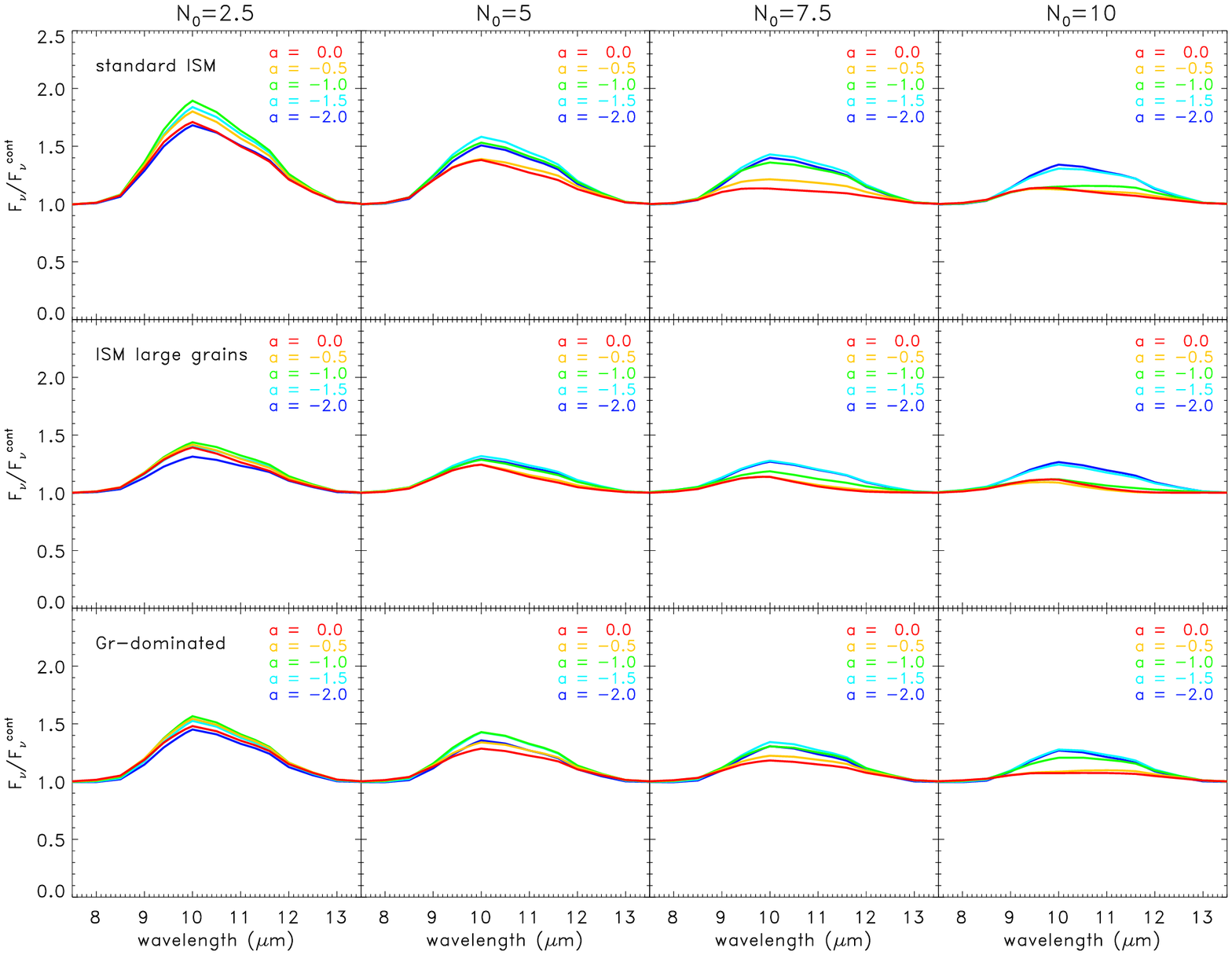}
\caption{Continuum-normalized $10\,\micron$ model silicate features in type-1 AGN (inclination angle $i=15\deg$). The top row show model models with the standard ISM composition, the middle row is for large grains, and the bottom row represents Gr-dominated dust. From left to right, the columns show an increasing mean number of clouds in the equatorial line-of-sight, $\No=2.5$, $5$, $7.5$, and $10$, respectively. In each panel, we plot SEDs of a representative random cloud distribution for radial power law indices $a=-0.5$ (red), $-1.0$ (yellow), $-1.5$ (green), and $-2.0$ (blue). Fixed parameters are $\Rout=150$, $\theta_0=45^\circ$, $\Rclo=0.035$, $b=1$, and $\taucl=50$.}\label{fig:tor_rad_Siem}
\end{figure*}

\begin{figure*}
\centering
\includegraphics[width=1.0\textwidth]{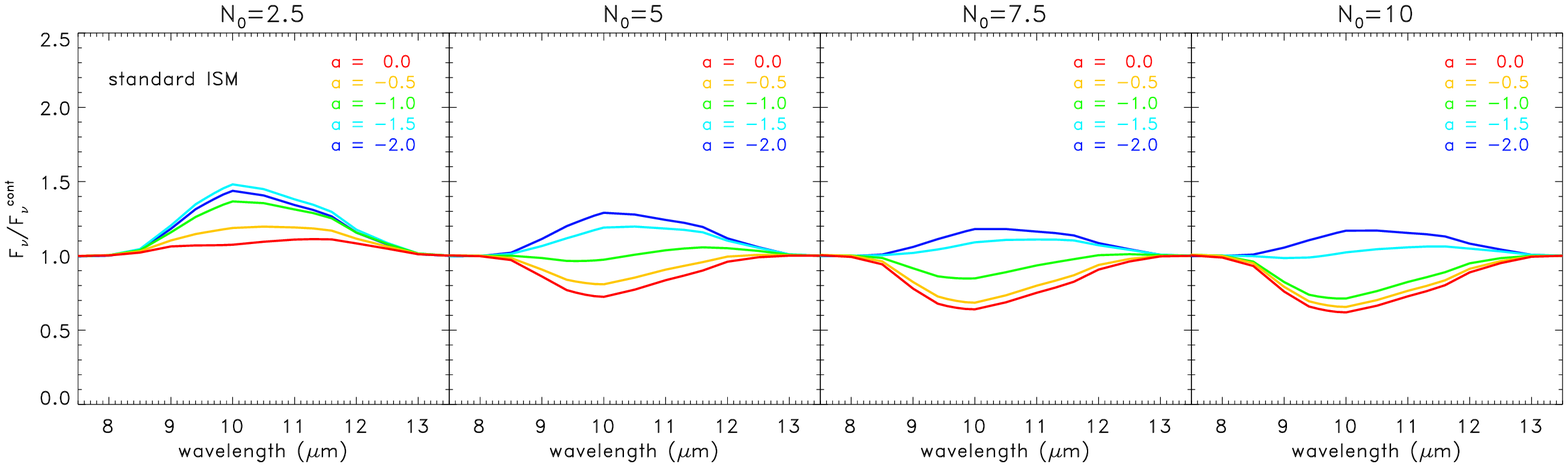}
\caption{Continuum-normalized $10\,\micron$ model silicate features in type-2 AGN (inclination angle $i=90\deg$) using the ISM standard dust composition. From left to right, the panels show an increasing mean number of clouds in the equatorial line-of-sight, $\No=2.5$, $5$, $7.5$, and $10$, respectively. In each panel, we plot SEDs of a representative random cloud distribution for radial power law indices $a=-0.5$ (red), $-1.0$ (yellow), $-1.5$ (green), and $-2.0$ (blue). Fixed parameters are $\Rout=150$, $\theta_0=45^\circ$, $\Rclo=0.035$, $b=1$, and $\taucl=50$.}\label{fig:tor_rad2_Si}
\end{figure*}

In the previous Sect.~\ref{sec:raddistr}, it has been shown that the effect of varying $\No$ on the overall SED is minor. On the other hand, the silicate feature in both type 1 and type 2 AGN changes from strong emission to weak emission or absorption with $\No$ increasing from 2.5 to 10.

In Figs.~\ref{fig:tor_rad_Siem} \& \ref{fig:tor_rad2_Si}, we show the $10\,\micron$ silicate features for different model parameters for type 1 and type 2 AGN respectively. In the case of type 1 AGN we show feature strengths for Ossenkopf ISM standard, large grain, and Gr-dominated dust mixes. To isolate the features, we followed the continuum spline fit procedure proposed by \citet{Sir08}. For that a cubic spline has been fitted to the regions at 5--7\,$\micron$, 14--14.5\,$\micron$, and 25-31.5\,$\micron$. The fitting result is considered representative for the ``continuum'' underlying the silicate features at 10 and 18\,$\micron$. While this method is applicable to Spitzer data, ground-based data usually have a much smaller wavelength coverage due to the atmospheric cutoffs \citep[e.g.][]{Hon10}. For such data we recommend the use of a linear fit to the 8.5 and 12.5\,$\micron$ fluxes as continnum representation. These wavelengths should be reasonably unaffected by the 10\,$\micron$ silicate feature -- at least at typically observed strengths \citep[e.g.][]{Mas06,Mas09,Hor08,Hon10}. As in the previous figures, we present increasing $\No$ from 2.5 to 10 in the columns from left to right. In each column, we color-coded the individual $a$-indices from $-0.0$ to $-2.0$ as in Fig.~\ref{fig:tor_rad_N}. From top to bottom, the rows show standard ISM, large grain, and Gr-dominated dust compositions. In general, the ISM standard configuration shows stronger emission features than the other dust configurations. 

Most of the change in the strength of the silicate feature happens when varying $\No$. Higher $\No$ corresponds to weaker emission features in type 1s. The same applies to type 2 AGN, however with a tendency of weaker emission features and slightly more pronounced absorption features than in type 1s. In addition to the influence of $\No$ on the feature strength, shallower dust distributions show less pronounced silicate emission features and stronger absorption features than steeper dust distributions. This is due to the interaction of obscuration in vertical direction (for larger $\No$, more clouds are present in vertical direction) and cloud distribution (flatter distributions are dominated by cooler dust emission).

In summary, while the overall SED depends only on one parameter (for a given dust distribution; as shown in the previous section), the exact strength of 10\,$\micron$ silicate feature is depending on at least $\No$ and $a$, with some emphasis on the former. Other parameters which may have an influence on the silicate feature are discussed in the next sections.

The standard ISM dust configuration with $\taucl$ around 50 seems to capture the essence of a typical torus SED including moderate silicate emission in type 1 and absorption in type 2 line-of-sights for a broad range of model parameters. We will, thus, use the ISM standard dust with Ossenkopf silicates in the following sections.

\subsubsection{The opening angle of the torus}\label{sec:theta}

\begin{figure*}
\centering
\includegraphics[width=0.9\textwidth]{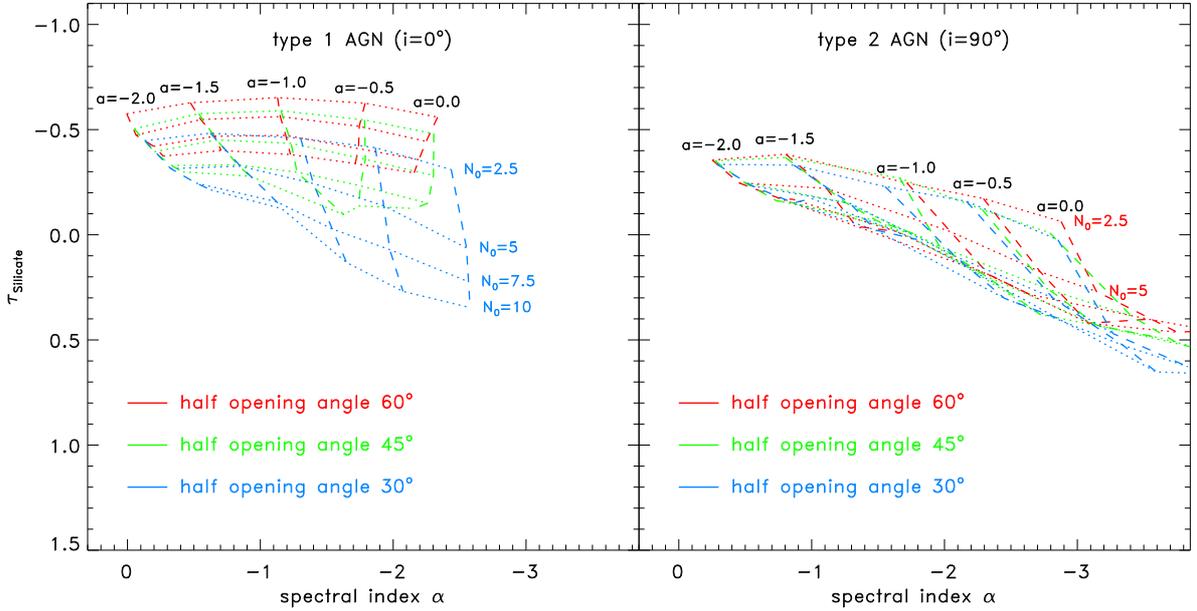}
\caption{Dependence of observed silicate-feature optical-depth $\tausi$ and mid-IR spectral index $\alpha$ on model parameters. The grids show how $\tausi$ and $\alpha$ change with $a$ (dotted lines; from $a=0.0$ to $-2.0$) and $\No$ (dashed lines; from $\No=2.5$ to 10) for different half-opening angles of the torus (red: 60$^\circ$, green: 45$^\circ$, blue: 30$^\circ$). \textit{Left:} Type 1 AGN with $i=0^\circ$; \textit{Right:} Type 2 AGN with $i=90^\circ$. The grid lines are results for one random arrangement of clouds for the given set of parameters.}\label{fig:theta0}
\end{figure*}

A potential source of uncertainty in determining $\No$ from the silicate feature (and maybe $a$ from the continuum SED) could be the geometrical thickness of the torus that we parametrized as the half-opening angle $\theta_0$. In Fig.~\ref{fig:theta0} we show how changing $\theta_0$ affects the mid-IR spectral index $\alpha$ and the silicate feature for standard ISM dust. For that we plotted the ``apparent'' optical depth $\tausi$ in the silicate feature as determined in the previous section versus the spectral index $\alpha$. The mid-IR spectral index $\alpha$ has been determined from the continuum spline fit as described in Sect.~\ref{sec:n0}. We then made a linear fit to the 7.0$-$8.5\,$\micron$ and 13.9$-$14.6\,$\micron$ region to determine the spectral slope in the $N$-band from which $\alpha$ has been calculated. Within Fig.~\ref{fig:theta0} we show the location of models ranging from $a=0.0$ to $-2.0$ and with $\No=2.5-10$ for $\theta_0=60^\circ$, 45$^\circ$, and 30$^\circ$. The left panel shows results for a pole-on view on the torus, the right panel represents a type-2-like edge-on view. Varying $\theta_0$ mainly affects the strength of the silicate feature in a way that emission features are weakened when the half-opening angle is smaller. The reason is that smaller $\theta_0$ results in more clouds along vertical line-of-sights which on one hand obscure the emission features from the torus mid-plane in type 1 AGN and on the other hand preferentially expose their cooler sides (= less strong emission features) to the observer. This effect is less pronounced for type 2 AGN where clouds affected by obscuration are those on the far-side of the torus which would be seen almost with their full hot emission side. The overall $\alpha$ and $\tausi$ varies less in type 2 line-of-sights when varying $\theta_0$.

In summary, varying $\theta_0$ alters primarily the silicate feature depth in type 1 cases while the spectral index is almost unaffected. Thus, using observables $\alpha$ and $\tausi$ leaves us with a degeneracy in model parameters $\No$ and $\theta_0$. However, sample studies showing type-1/type-2 ratios of 1:1 to 1:3 \citep[e.g.][]{Mai95,Lac04,Mar06} and observed narrow-line region opening cones around 90$^\circ$ in many nearby objects favor $\theta_0\sim40-50^\circ$. For individual objects, $\theta_0$ may be smaller or larger \citep[e.g.][]{Mas09} than this average. When considering, however, a larger sample of AGN, the use of $\theta_0=45^\circ$ as the sample average is well justified to break the $\No$-$\theta_0$-degeneracy and get an idea what range of $\No$ is covered by the AGN in the sample. On the other hand, the uncertainty in the individual clouds' optical depths remains in such an analysis unless independent evidence narrows down $\taucl$ (e.g. general agreement of SEDs with unification scheme, IR visibilities, Hydrogen column densities, theoretical considerations). Thus it seems more viable to give a sample $\No$-$\taucl$-range or providing the $\No$ constraint under the assumption of a certain $\taucl$.

\subsubsection{The outer radius of the torus}\label{sec:rout}

When modeling the IR SEDs, there is often a question arising: How far outward does the torus extend in the model? Since our model defines $\Rout$ as a parameter, one might think that the $\Rout$ value would be the correct answer. However, it is not all that simple. Due to the decrease of cloud temperature with distance and the corresponding decrease of total intensity, clouds at different distances from the AGN contribute different fractions to the total torus flux at a given wavelength. In general, clouds at the innermost torus region contribute most of the near-IR light while clouds at larger radii are the dominant source of the mid-IR emission. Being more quantitative we can approximate
\begin{equation}\label{eq:radeq}
\frac{r}{\rsub} = \left(\frac{T}{T_\mathrm{sub}}\right)^{-2\ldots-2.8}
\end{equation}
for dust grains in radiative equilibrium \citep[e.g.][]{Bar87}, where $T_\mathrm{sub}$ is the dust sublimation temperature ($\sim$1500\,K) and $T$ is the dust temperature at distance $r$ from the AGN. The range given for the exponent covers the range from black body grains (large, graphite) to typical ISM dust \citep[see also][]{Bar87}. Based on this equation and Wien's law, it is possible to estimate the maximum radius which contributes to the torus emission at a given wavelength. Thus, the maximum size of the emission is depending on the observed wavelength. On the other hand, the exact shape of the SED (and also the observed size of the emission region; see Sect.~\ref{sec:intf}) depends on the dust distribution (see previous sections), so that eq.~(\ref{eq:radeq}) provides only a rough estimate. However, choosing $\Rout$ smaller than the size corresponding to the temperature (= wavelength) of interest may result in an artificial cut-off in the brightness distribution of the models. For example, when selecting $\Rout=25$, the cut-off occurs at temperatures of $T\sim475$\,K for ISM dust which corresponds to wavelengths around 8\,$\micron$.

The torus brightness distribution does not only depend on the dust temperature (or source function of clouds) but also on the dust distribution as shown in Sect.~\ref{sec:modparams}. Steep dust distributions have most of their dust at small radii so that they are mostly not sensitive to $\Rout$. However, the actual choice of $\Rout$ will become important for the mid-IR SED once the dust distribution is very flat ($a\ga-1$) or even inverted ($a\ga0$)\,\footnote{$a=-1$ seems to be a borderline case. Although the integration of the brightness profile does not formally converge, obscuration effects will reduce the dependency on $\Rout$ if properly selected.}. In this case the integration of the brightness profile will not converge for $\Rout\rightarrow\infty$ and the result depends on $\Rout$ (see eq.~(\ref{eq:master_simple})). It is, thus, important to select $\Rout$ properly to avoid artificial cut-offs in intermediate cases. Moreover for quite extended distributions, such as the presented $a=0.0$ case, $\Rout$ has to be considered as a full model parameter.

Recent IR interferometric observations and $K$-band reverberation mapping measurements suggest that the torus brightness distributions are quite extended at least up to $\sim100\times\rsub$ \citep{Kis09a,Kis09b,Hon10} in contrast to the discussion in \citet{Nen08b}. For a better constraint on the ``real'' $\Rout$ the use of ALMA at even longer wavelengths will be crucial. However, it may be difficult to set a strict limit of the torus (= AGN-heated dusty region) with respect to its galactic environement (= star or starformation-heated).

\subsubsection{Using SEDs to determine the radial power-law index $\alpha$}\label{sec:sed_sum}

\begin{figure}
\centering
\includegraphics[width=0.5\textwidth]{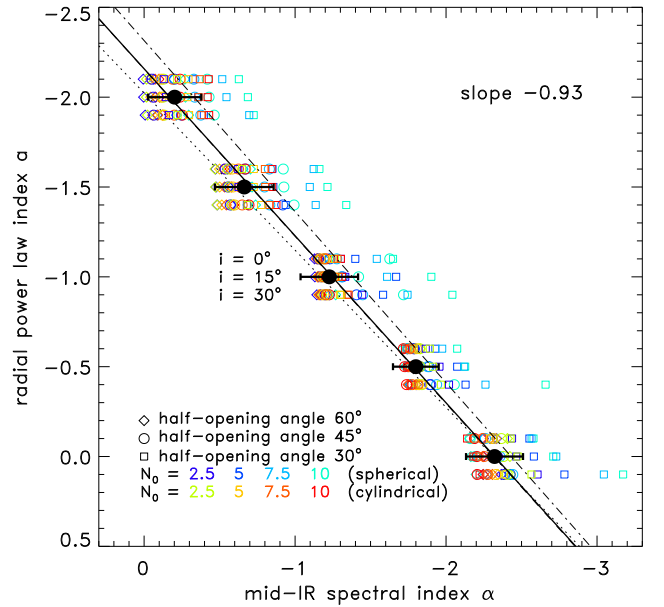}
\caption{Relation between the mid-IR spectral index $\alpha$ and the radial dust distribution power-law index $a$ in type 1 AGN. Model points are shown for all combinations of $\No=2.5$, $5$, $7.5$, and $10$ (color-coded symbols) and half-opening angle $30^\circ$ (squares), $45^\circ$ (circles), and $60^\circ$ (diamonds) for both spherical (bluish-colored symbols) and cylindrical (reddish-colored symbols) vertical distributions. In order to cover a range of type-1 inclinations, results for $i=0^\circ$ (face-on; upward offset), $i=15^\circ$ (nominal $a$ value) and $i=30^\circ$ (downward offset) are presented for all $a$-values. The black-filled circles with errors denote the nominal $\alpha$-value for each $a$ and one standard deviation. The nominal fit using all values (solid black line) has a slope of $-0.93$ and a Spearman rank correlation of $-0.97$. Fits using only results with half-opening angle $30^\circ$ (dash-dotted line), $45^\circ$ (dashed line; almost coincident with the solid line), and $60^\circ$ (dotted line) are also shown for illustration.}\label{fig:alpha_a}
\end{figure}

Summarizing the subsection on torus SEDs, we showed that the slope in the mid-IR, represented by $\alpha$, is strongly related to the radial dust distribution in the torus. This seems to be the case despite several other parameters which potentially influence the torus emission. It is, however, consistent with expectations based on theoretical considerations (see Sect.~\ref{sec:modparams}).

In Fig.~\ref{fig:alpha_a} we present a more quantitative approach to the relation between $\alpha$ and $a$ in type 1 AGN. For each model $a$ parameter, we varied $\theta_0$ (different symbols) and $\No$ (color-coded), and plotted the resulting $\alpha$ versus $a$. In addition we assess the influence of torus inclination in type 1 AGN by varying $i$ from 0$^\circ$ to 30$^\circ$ (offsetted in the plot). To be even more general we also included simulations for both the spherical (bluish colors) and cylindrical (reddish colors) vertical distributions (see Sect.~\ref{sec:modparams}). This results in 72 modeled $\alpha$ values for each $a$. The black filled circles with errors note the mean of these models for each $a$ and the standard deviation around the mean value\footnote{For convenience we calculated the standard deviation from all model $\alpha$ values for each $a$ value, although the distribution of $\alpha$'s around the mean is non-Gaussian, non-symmetric.}. These mean values have been fitted by a straight line with slope $-0.93$, and the overall correlation of $a$ vs. $\alpha$ has a Spearman rank of $-0.97$. The main outliers from the fit are geometrically very thick ($\theta_0=30^\circ$), higher inclination ($i=30^\circ$) cases with high $\No$ using the spherical distribution, while the general relation is quite tight. In fact the slope of the fit does not change too much when fixing $\theta_0$ (see lines for $\theta_0=30^\circ$ (dashe-dotted), 45$^\circ$ (dashed; almost coincidental with solid line), and 60$^\circ$ (dotted)), ranging approximately from $-0.87$ to $-0.95$. The nominal fit relation for the mean values including errors is 
\begin{equation}
a = (-2.16\pm0.16) - (0.93\pm0.11)\cdot\alpha\,\,. 
\end{equation}
Within the range of errors illustrated in Fig.~\ref{fig:alpha_a}, this relation may be used to extract information about the dust distribution from mid-IR spectra.

\subsection{Interferometry of dust tori}\label{sec:intf}

Due to the mas-spatial scales involved in torus observations, it is difficult to spatially resolve the torus. The most promising tool for this task is IR interferometry, which was already successfully used for a number of nearby objects \citep{Swa03,Wit04,Wei04,Jaf04,Pon06,Mei07,Tri07,Bec08,Rab09,Tri09,Bur09,Kis09b}. However, except for reconstructed images of NGC~1068 obtained from Speckle interferometry \citep{Wei04}, long-baseline interferometry usually provides spatial information coded as Fourier space observables, i.e. phase and visibility information for a given (projected) baseline and position angle. While the object's phase is important for image reconstruction, the visibility contains the fundamental spatial information about the object (size, elongation). This information can be used as complementary to total flux SEDs, in particular if for a given wavelength both total and correlated fluxes are available.

The figures in the following subsections are simulated for an AGN at 15\,Mpc distance with an inner radius $\rsub=0.025\,\pc$. They can be used as predictions for any AGN by re-scaling the $x$-axis baseline scale according to
\begin{equation}
BL_\mathrm{AGN} = BL_\mathrm{model} \cdot \left(\frac{D_{A;15}}{r_{0.025}}\right)
\end{equation}
where $D_{A;15}$ is the AGN's angular-diameter distance in units of 15\,Mpc and $r_{0.025}$ is the reverberation/sublimation radius in units of 0.025\,pc.

\subsubsection{The baseline-dependence of the visibility}\label{sec:blvis}

\begin{figure*}
\centering
\includegraphics[width=1.0\textwidth]{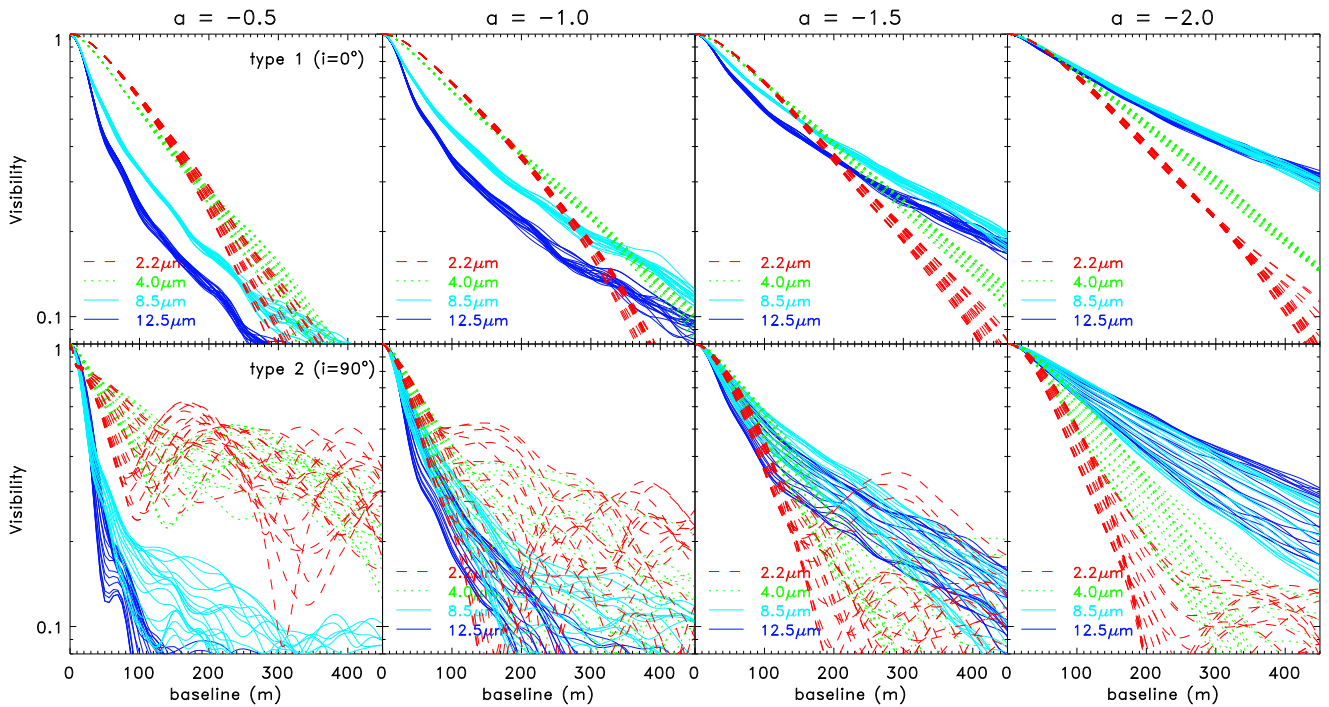}
\caption{Dependence of the model visibility on the baseline for an AGN at 15\,Mpc distance and a sublimation radius (= near-IR reverberation radius) of 0.025\,pc. Each column represents simulation results for a different radial power law index $a=-0.5$, $-1.0$, $-1.5$, and $-2.0$, respectively, (top: type 1 AGN at $i=0\deg$; bottom: type 2 AGN at $i=90\deg$) with constant $\No=7.5$ and $\Rout=150$. In each panel, we show visibility curves at $2.2$ (red-dashed lines), $4.0$ (green-dotted lines), $8.5$ (light-blue solid lines), and $12.5\,\micron$ (dark-blue solid lines). The different lines per wavelength reflect the position-angle dependence of the visibility in steps of $10^\circ$ and represent one particular random distribution of clouds.}\label{fig:tor_vis_rad}
\end{figure*}

As shown in literature, adding interferometric information is very constraining for modeling the dust distribution of the torus. In particular, small-scale position-angle-dependent variations of the visibility for a given baseline and wavelength seem to be clear evidence for the clumpy structure of the torus \citep{Hon06,Sch08}. However, interferometry constrains quite direct information about the radial intensity distribution of the torus emission: When observing almost face-on-projected type 1 AGN, no significant position-angle change of the visibility is expected. Thus, if data for different baseline lengths (corresponding to different resolution scales) is available, we have a direct and model-independent access to the surface brightness distribution $I_s$ of the torus \citep[see also][]{Kis09a}. As shown in Sect.~\ref{sec:modparams}, $I_s\propto S_\nu\sigs$ primarily depends on $a$. By modeling the baseline- and wavelength-dependence of the visibility, we will be able to tightly constrain the radial dust distribution, mostly independent of the SED.

In Fig.~\ref{fig:tor_vis_rad}, we show the baseline-dependence of the visibility for four different wavelengths. The wavelengths have been selected to be in line with current and future interferometric instruments at the VLT and the Keck interferometers: the $K$-band at $2.2\,\micron$ (e.g. VLTI/AMBER), $4.0\,\micron$ in the $L/M$-band region (e.g. KI or future VLTI/MATISSE), and $8.5$ and $12.5\,\micron$ in the $N$-band (e.g. VLTI/MIDI). For the simulations, we use fixed $\No=7.5$ and $\Rout=150$ and present results for different radial power law indices $a=-0.5$, $-1.0$, $-1.5$, and $-2.0$. The top and bottom rows in Fig.~\ref{fig:tor_vis_rad} show $i=0\deg$ and $i=90\deg$, respectively. For all wavelengths the variation with position angle in steps of 10$^\circ$ is shown. Each column in Fig.~\ref{fig:tor_vis_rad} represents one particular random arrangement of clouds for the given set of torus parameters.

We will first concentrate on the type 1 case where the observed visibilities can be directly used to constrain the dust distribution. For the shown baseline lengths from 0$-$450\,m, the $K$-band visibilities in the visibility range $\ga$0.3 have an overall similar shape with only little dependence on $a$. For instance, the $K$-band visibility at 150\,m is 0.5 for $a=-2.0$ and 0.4 for $a=-0.5$. This similarity can be understood since most of the $K$-band emission is originating from the hot dust which is located in the innermost region of the torus, regardless of the actual torus-internal obscuration or dust distribution.

\begin{figure*}
\centering
\includegraphics[width=1.0\textwidth]{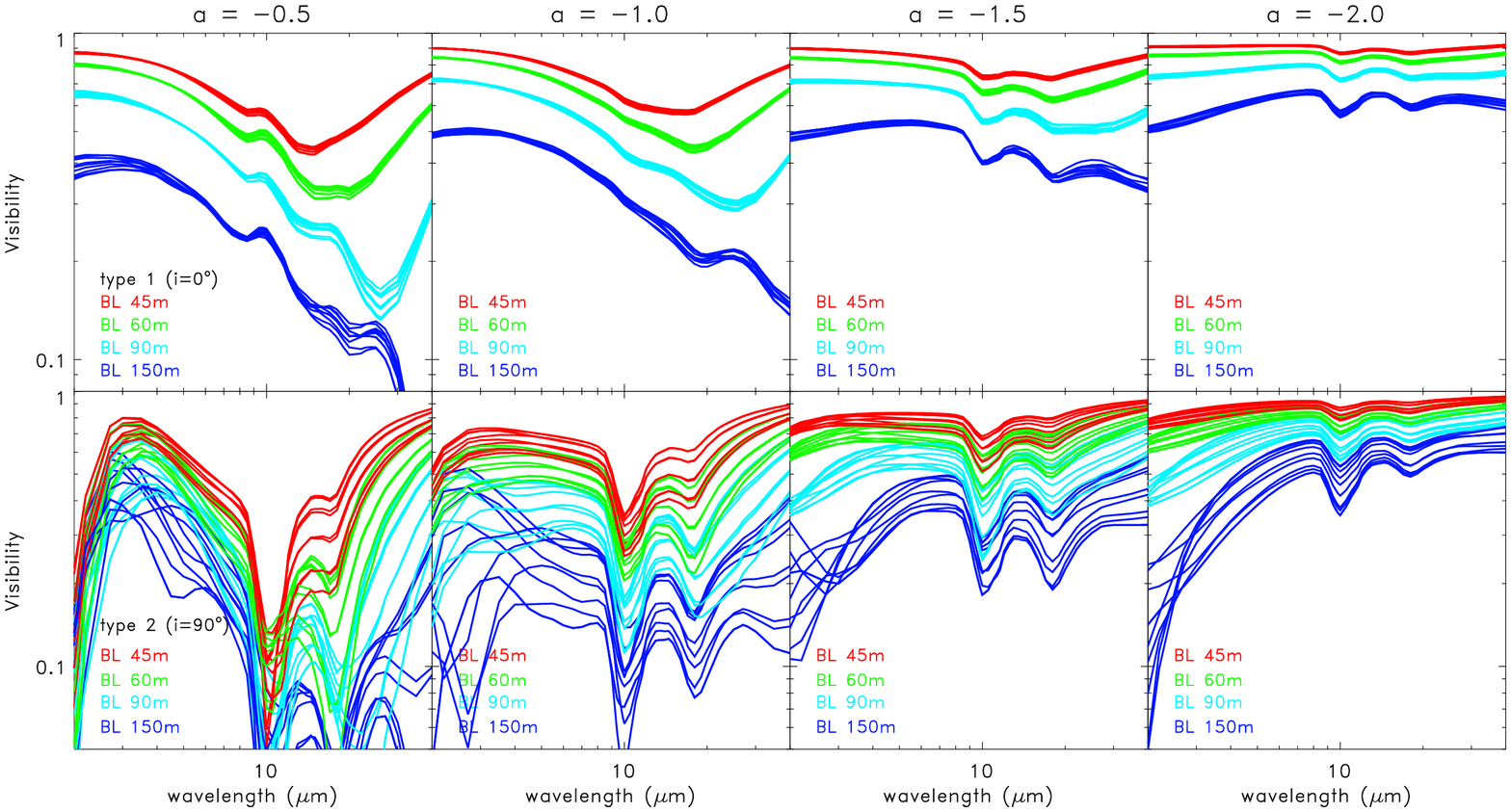}
\caption{Wavelength-dependence of the model visibilities for different baselines for an AGN at 15\,Mpc distance and a sublimation radius (= near-IR reverberation radius) of 0.025\,pc. Each row represents simulation results of one random arrangement of clouds for different radial power law index $a$ (top: type 1 AGN at $i=0\deg$; bottom: type 2 AGN at $i=90\deg$) with fixed $\No=7.5$ and $\Rout=150$. In each panel, we show visibility curves for baseline lengths 45\,m (red lines), 60\,m (green lines), 90\,m (light-blue lines) and 150\,m (dark-blue lines). The different lines per wavelength reflect the position-angle dependence of the visibility.}\label{fig:tor_vis_spec}
\end{figure*}

In a smooth dust distribution we would expect no position-angle dependence of the visibility in a type 1 torus. The fact that we see this PA variations is an evidence for clumpiness. The strength of these variations is depending on wavelength. Individual dust clouds are optically thick in the $K$-band (see Sect.~\ref{sec:dustcloud}), so that the visibilities can vary between different PAs due to the actual clouds' placements and resulting torus-internal obscuration effects. At longer wavelengths, the individual clouds are optically thin and the variations with position angle becomes smaller. The variations still seen at these wavelengths are coming from a combination of ``clumping of clumps'' in the torus, resulting in both stronger local emission and absorption, or from ``holes'' in the distribution. From this analysis, we can draw an interesting first conclusion: The strength of the position-angle dependence of the visibility in type 1 AGN is related to properties of the clouds. This concerns, in particular, the optical depth of the clouds which is related to the cloud dust mass or geometrical properties and the dust composition. In addition, statistical effects will play a role: the same overall torus parameters ($a$, $\No$, $\theta_0$), can be kept constant when increasing the cloud sizes $\Rcl$ by decreasing the total number of clouds in the torus (see Sect.~\ref{sec:modparams}). Less number of clouds increase statistical effects in the distribution (i.e. holes, obscuration along certain line-of-sights) so that the ``clumpiness variations'' increase.

As previously mentioned, the $K$-band visibility curves at visibilities $\ga$0.3 are very similar when comparing different $a$, so that it can be used as a reference for the other wavelengths. As can be seen in Fig.~\ref{fig:tor_vis_rad}, the baseline-dependence of the visibility changes with $a$ for longer wavelengths: The $N$-band visibilities increase from $V_{12.5\,\micron}=0.15$ to $0.65$ at 150\,m when $a$ changes from $-0.5$ to $-2.0$. The basis of this behavior is the spatial regions and temperatures involved in emitting light at the given wavelengths. At short wavelengths, only hot dust contributes to the emission and no light will be emitted from cool dust. This makes the emission region small, no matter how the dust is distributed. In a simplified picture of a type 1 AGN, we expect that only a very small ring is seen, approximately with the size corresponding to the equilibrium temperature peaking at this wavelengths. At longer wavelengths both hot and cooler clouds contribute to the flux. Thus, the emission region appears more extended. The actual extension, however, depends on how the dust is radially distributed. If the density is steeply decreasing outwards, the size is still small and comparable to only hot emission. If the density distribution is flat, larger distances contribute significantly and the size is much more extended than for hot emission.

The described principles are also valid for type 2 objects. However, in type 2 AGN obscuration within the torus becomes more important. In the bottom row of Fig.~\ref{fig:tor_vis_rad}, we show the baseline-dependence of the visibility for an inclination angle $i=90\deg$, i.e. edge-on as in type 2 AGN. Since clouds at larger radii along our line-of-sight obscure clouds at smaller radii (where most of the hot emission is originating), the brightness contrast between regions close to the AGN and further away decreases, so that the tori appear more extended. Thus, the visibilities for a given baseline and wavelengths are generally smaller than in the type 1 case. Since obscuration plays an important role, the small-scale variations of the visibility are stronger. In addition and as expected, the tori show elongation along the mid-plane which is responsible for a large fraction of the position-angle dependence of the visibility shown in Fig.~\ref{fig:tor_vis_rad} (bottom). Similar to the type-1 case, the $K$-band visibility shows the strongest small-scale visibility variation because the clouds are optically thick. The essential dependence of the visibility curves on wavelength are also comparable to the type 1 case: Longer wavelengths show a stronger change from steep to flat for $a$ changing from $-0.5$ to $-2.0$. It is interesting to note that at $a=-2.0$, the type 2 AGN visibilities are very similar to type 1 AGN. This can be understood when considering that in the steep cases, most of the dust is located within few sublimation radii. Thus, the overall emission is dominated by hot clouds without obscuration from larger scales. 

In summary, as illustrated in Fig.~\ref{fig:tor_vis_rad}, the baseline-dependence of the visibility in type 1 AGN directly traces the radial dust distribution in the torus: If the dust distribution is steep ($a=-1.5$ or $-2.0$), the baseline-dependence of the visibility is comparably flat, and vice versa, and the effect becomes stronger for longer wavelengths. In principle, this also applies to type 2 AGN, but obscuration effects and small-scale visibility variations make the $a$-dependence of the visibility less obvious. Although we only show results for $\No=7.5$, we note that these conclusions are mostly independent of $\No$ for the range of $\No=2.5$ to 10 that we investigate in this paper. 

\subsubsection{The wavelength-dependence of the visibility}\label{sec:specvis}

\begin{figure*}
\centering
\includegraphics[width=1.0\textwidth]{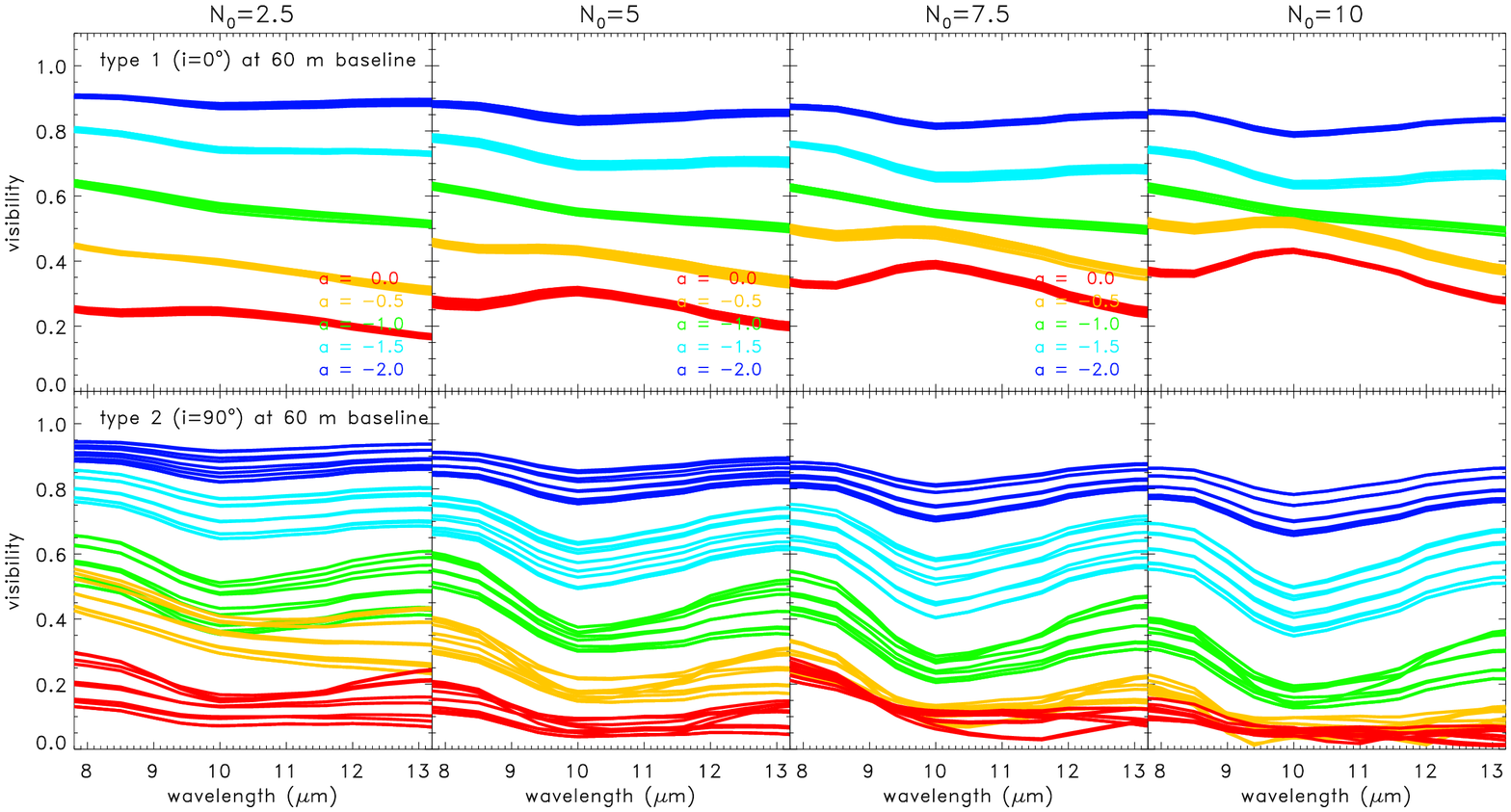}
\caption{Wavelength-dependence of the model visibilities in the MIDI wavelength range from $8-13\,\micron$ for an AGN at 15\,Mpc distance and a sublimation radius (= near-IR reverberation radius) of 0.025\,pc at a baseline of 60\,m. Each column represents simulation results of one random arrangement of clouds for different $\No$ ranging from 2.5 (left) to 10 (right). In each panel we show models for radial power law indices $a=0.0$ (red), $-0.5$ (orange), $-1.0$ (green), $-1.5$ (light blue), and $-2.0$ (dark blue). The top row represents a type 1 AGN at $i=0\deg$, the bottom row a 2 AGN at $i=90\deg$. The different lines per $a$ model reflect the position-angle dependence of the visibility.}\label{fig:tor_vis_spec_N}
\end{figure*}

In Fig.~\ref{fig:tor_vis_rad}, we can see that for the ``extended'' tori with $a=-0.5$ and $-1.0$, the visibility curves at wavelengths longer than $2.2\,\micron$ are below the $K$-band visibility curve. In other words, the wavelength-dependent size of the torus grows faster than the wavelength-dependent resolution decreases for a given baseline. On the other hand, the ``compact'' tori with $a=-1.5$ and $-2.0$ show the opposite behavior, i.e. the long wavelength visibility curves are close to or above the $K$-band curve for visibilities below 0.5. Of course, this reflects the fact that most of the dust is confined to the inner torus region, making the apparent size small at all wavelengths. While this finding is true for both type 1 and type 2 orientations, it is much more obvious in face-on geometries where obscuration effects do not play a significant role.

Instead of looking at the baseline-dependent visibility at different wavelengths, we can also compare the wavelength-dependence of the visibility for given baselines. In Fig.~\ref{fig:tor_vis_spec} we show visibilities in the 2 to 40\,$\micron$ range for both type 1 (top row) and type 2 (bottom row) orientations of the torus. As in the previous figure, we plot curves for $a=-0.5$, $-1.0$, $-1.5$, and $-2.0$ (from left to right column), $\No=7.5$, $\Rout=150$, and $\theta_0=45^\circ$ are fixed. Each panel contains simulations for baseline lengths of 45\,m (red), 60\,m (green), 90\,m (light blue), and 150\,m (dark blue). For each configuration, various lines illustrate the position-angle dependence. The visibility usually drops strongly with increasing wavelength up to the mid-IR in the extended cases ($a=-0.5$ and $-1.0$) for both type 1s and type 2s. The turn-around to increasing visibility is depending on baseline and occurs at $\sim14\,\micron$ for 45\,m, $\sim18\,\micron$ at 60\,m, and $\sim27\,\micron$ at 90\,m. This trend is less obvious in the type 2 cases where the 10 and 18\,$\micron$ silicate features are seen in the visibilities. The overall increase in turn-around wavelength for increasing baseline length goes along with the increase in spatial resolution. From this we conclude that up to the turn-around wavelength, the size of the corresponding emission region increases faster than the loss of wavelength-dependent spatial resolution for a given baseline. On the other hand longer wavelengths would correspond to cooler dust in the outer regions of the torus where both obscuration effects suppress direct AGN heating and our model space ends ($\Rout=150$ corresponds to cloud temperatures of $70-180$\,K or $>16\,\micron$). Thus, the size of the emission region cannot grow any further. In the case of compact dust distributions ($a=-1.5$ and $-2.0$) all clouds are confined to small radii so that there are almost no clouds contributing from cooler regions. As a result the size of the torus is very similar at all wavelengths, so that the loss of resolution with increasing wavelength for a fixed baseline is dominating the trends of visibilities.

The difference of the wavelength-dependent visibility for different selections of $a$ opens an interesting possibility to narrowly constrain the radial distribution of the dust, at least in type 1 AGN: By obtaining interferometric visibilities for an AGN at 2 different wavelengths for the same baseline length -- preferably one visibility point in the near-IR, one in the mid-IR -- it is possible to constrain the radial dust distribution index $a$ or, at least, characterize the radial dust distribution as ``extended'' ($a \ge -1.0$) or ``compact'' ($a \le -1.5$). Such experiments can, for example, be done using the UT telescope baseline configurations at the VLTI in combination with the AMBER and the MIDI instruments. On the other hand, since MIDI already provides some wavelength coverage from 8 to 13\,$\micron$, using MIDI alone may already give some hint. In Fig.~\ref{fig:tor_vis_spec_N} we show model visibilities for $\No=2.5$ to 10 for $a=0.0$ to $-2.0$ for a baseline of 60\,m. $\theta_0=45^\circ$ and $\Rout=150$ have been fixed. The strongest difference can be seen between the extreme cases in type 1s: While overall trend in visibilities for $a=-2.0$ is flat for all $\No$ (same visibilities at 8 and 13\,$\micron$), the shallow distributions (e.g. $a=-0.5$) show some clear trend of decreasing visibility in the same wavelength range.

As shown in Fig.~\ref{fig:tor_vis_spec_N} the silicate feature appears in the visibilities. But unlike in the SEDs, it is very shallow for small $\No$ and seems in ``absorption'' for steep dust distributions and in ``emission'' for flat ones in case of type 1s. In type 2s it is always in ``absorption''. This tells us that the feature-emitting region is more extended with respect to the continuum in steep distributions and type 2s while it is more compact in shallow distributions. Given the dependence on $\No$ (larger $\No$ increases both effects), this difference can be ascribed to torus-internal obscuration effects. Since optical depth effects along the line-of-sight are responsible for the exact strength of the feature, we expect that, aside from $a$ and $\No$, other parameters will also have an influence, such as $\theta_0$ (as shown in Sect.~\ref{sec:theta} for the SED), the dust chemistry and the optical depth of the individual clouds. Interestingly, at around $a=-1$ the type 1 visibilities do not show the feature which might indicate canceling out of some of the parameter dependencies. The presence of a silicate ``absorption'' feature in the visibilities of type 2 geometries have been claimed to be a unique feature to clumpy torus models. Smooth dust distributions are allegedly not showing this feature in the visibility spectrum \citep{Jaf04,Sch05}. On the other hand, a silicate absorption feature is prominently displayed in interferometric data of NGC~1068 \citep{Jaf04,Pon06,Rab09}. Here, we show that clumpy models can show a variety of feature appearances in the visibilities for both type 1 and type 2 AGN.

\subsubsection{The constraining power of IR interferometry on torus properties}\label{sec:intf_sum}

As a summary of the model results for IR interferometry we emphasize the constraining power that IR interferometry has when combined with proper modeling. Here the radial dust distribution is particularly accessible in type 1 AGN. This has been demonstrated recently by \citet{Kis09b} on a set of interferometric data of type 1 AGN and in our paper I for NGC~3783 \citep{Hon10}. The main reason for this constraining power is that IR interferometry can measure wavelength-dependent sizes of the continuum emission region which corresponds to the temperature distribution inside the torus and is, in turn, determined by the dust distribution and the source function of the clouds (see eq.~\ref{eq:master_simple}). Since we showed in Sect.~\ref{sec:sed_sum} that the mid-IR continuum slope is also determined by the radial dust distribution, a correlation between mid-IR spectral index $\alpha$ and the compactness in interferometric observations can be expected (compact sources have bluer $\alpha$).

The silicate feature observed in visibilities has less constraining power on the models. The reason is that there exists some degeneracy in parameters which control the strength of the feature. We have seen in Fig.~\ref{fig:tor_vis_spec_N} that the ``absorption'' and ``emission'' feature in type 1 AGN change with $a$ contrary to what has been seen in the SEDs (see Fig.~\ref{fig:tor_rad_Siem}), which essentially tells us how the size of the emission regions of the feature change. The actual strength, however, is still influenced by $\No$ (see Fig.~\ref{fig:tor_vis_spec_N}) and $\theta_0$ (as in Fig.~\ref{fig:theta0}), and probably also by the optical depth and chemical composition of the individual dust clouds. It may be still possible that simulatneous modeling of SED and interferometry can break some of the degeneracy and provide information about the dust cloud properties (optical depth) using the strength of the ``clumpiness variations''.

\section{Summary and conclusion}\label{sec:sumconc}

We presented modeling results from our upgraded radiative transfer model of 3D clumpy tori and described how observations can constrain some of the parameters used in the model. The upgrade concerns the handling of the diffuse raditation field within the torus for which we presented a statistical approximation. We outlined our modeling strategy and introduced the mathematical foundation. The model parameters have been connected to properties of the torus, such as the volume filling factor or the cloud number density. We also derived a simplified type 1 torus model from the full and more complex model. This simplified model has been used to illustrate which geometric parameters have the strongest impact on the torus emission, which are the radial dust cloud distribution power-law index $a$, the mean number of clouds along a radial equatorial line-of-sight $\No$, and the torus height described either as the scale height $h$ or the half-opening angle $\theta_0$. We also noted that the cloud radii $\Rcl$ do not have a significant impact on the torus simulation results, while the cloud dust properties ($\tauV$ and dust composition) will probably be more important. 

In the next step, we presented results of Monte Carlo simulations of dust clouds using different dust compositions. We used standard ISM compositions with \citet{Dra03} and \citet{Oss92} silicates, respectively, and compared the cloud SEDs to simulations with a large grain composition and Graphite-dominated dust. The differences between the cloud SEDs from different dust compositions are an effect of the optical depth in the infrared and the different (Planck mean) opacities of each dust distribution which result in a slightly different temperature gradient with distance from the AGN.

By using the pre-calculated cloud SEDs and images, we simulated torus SEDs and studied how different parameters affect the model SEDs. The SEDs are presented in a suitable way to be scaled to observations. Our main conclusions are:
\begin{list}{$\bullet$}{}
\item The dust clouds with standard ISM dust seem to provide reasonable torus SEDs which seemingly capture the essence of recently observed IR characteristics of AGN (e.g. moderate 10\,$\micron$ silicate emission and absorption features). While not showing explicit dependence of the SEDs on the optical depth of the individual clouds, $\taucl=30-80$ for ISM standard dust is probably the range that provides good agreement with observations.
\item The torus SEDs, in particular in type 1 cases, are quite sensitive to the selection of $a$. Distributing more dust clouds at larger distances (e.g. $a=-0.5$ or $-1.0$) results in redder mid-IR colors. We showed that $a$ is the main parameters that determines the mid-IR spectral index $\alpha$ with only little dependence on $\No$ or $\theta_0$. Therefore it is possible to use suitable mid-IR data to directly constrain the radial distribution of the dust in the torus. A formal correlation function between $\alpha$ and $a$ has been extracted from the models (see Sects.~\ref{sec:raddistr} \& \ref{sec:sed_sum} and Fig.~\ref{fig:alpha_a}).
\item The appearance of the silicate feature -- either in absorption or emission -- is not only viewing-angle (inclination) dependent, as could be naively concluded, but depends on the radial and vertical distribution of the dust (i.e. $a$ and $\theta_0$ or $h$), the average number of clouds along an equatorial line of sight ($\No$) and possibly on $\taucl$ and the dust composition (see Sects.~\ref{sec:n0} \& \ref{sec:theta}). In fact, all the cloud and torus parameters influence the exact shape and depth of the silicate feature towards more absorption or emission and it may be difficult constraining individual parameters only based on silicate feature observations. That being said it is, however, crucial that parameters fitting the overall SEDs of observations are also in agreement with the silicate feature strength. We showed that compact dust distributions ($a=-1.5$ and $-2.0$) with $\taucl=50$ do not produce any silicate absorption features for $\No\le10$ in type 2 line-of-sights. Moreover, the silicate emission features in type 1 AGN become more pronounced with decreasing $\No$ and/or more compact $a$. From the commonly seen weak silicate emission features in type 1 AGN and silicate absorption features in type 2 AGN, we would expect that, in general, AGN dust tori should have, on average, rather extended dust distributions with $\No\ga5$. 
\item We present an easy way to compare torus models to observations. For that we plot the apparent silicate feature strength $\tausi$ against the mid-IR spectral index $\alpha$ which are the main observational properties of torus spectra (see Sect.~\ref{sec:theta}). On top of this, we overplot the parameter range that is covered by the models. Neglecting inclination effects, $\alpha$ is almost exclusively depending on $a$ while $\tausi$ is influenced by $\No$, $\theta_0$, and the cloud dust properties ($\tauV$ and the dust composition). If we select a separate sample of type 1 and type 2 AGN (i.e. having average inclinations of $i\sim30^\circ$ and $i\sim75^\circ$ respectively) and consider sample average type-1/type-2 ratios (leading to a an average $\theta_0\sim40-50^\circ$) it should be possible to provide some sample average constraint on a combined $\No$-$\taucl$. Considering both 10$\,\micron$ \textit{and} 18$\,\micron$ silicate features might help breaking some of the remaining degeneracy \citep{Tho09}. The principle of this approach has been introduced in paper I of this series \citep{Hon10}.
\item When modeling near- to mid-IR data, the outer radius does not have a significant influence on the near- and mid-IR emission for steeper dust distribution than $a\la-1$. In these cases the majority of the dust is confined to distances $\ll\Rout$. On the other hand, for $a\ga-1$ the dust distribution function does not converge for $\Rout\rightarrow\infty$, so that $\Rout$ influences the peak of the overall torus emission. It is therefore important to properly select $\Rout$ in all cases to avoid artificial cutoffs in the brightness distribution of the targeted wavelength range (see Sect~\ref{sec:rout}).
\end{list}

From torus model images, we calculated visibilities to simulate observations with interferometers. In addition to the SEDs, these data provide spatial information at different wavelengths and at different scales depending on the baseline length. We conclude:
\begin{list}{$\bullet$}{}
\item In type 1 AGN, the baseline-dependence of the visibility directly traces the radial distribution of the dust (here: the power law index $a$) without significant influence from other parameters (see Sect.~\ref{sec:blvis}). The general principle is also valid for type 2 AGN, but obscuration and geometric effects result in significant dependence of the visibility on position angle.
\item By comparing baseline-dependent visibilities at two different wavelengths (preferably one in the near-IR, one in the mid-IR) for a given baseline, one can directly conclude if the dust distribution is compact ($a\ge-1.5$) or extended ($a\le-1.0$). This can be done for both type 1 and type 2 AGN, although type 2s are less constraining due to obscuration effects (see Sects.~\ref{sec:blvis} \& \ref{sec:specvis}).
\item Small-scale position-angle variations of the visibility in type 1 AGN are usually associated with clumpiness of the torus (see Sect.~\ref{sec:blvis}). The strength of these variations is controlled by (1) the optical depth $\taucl$ of the cloud and, thus, also the dust composition (they are strongest if $\tau_\lambda\gg1$ at the observed wavelength $\lambda$), (2) the cloud size $\Rcl$, and (3) the number of clouds in the torus (i.e. statistics which are related to the choice of $\No$). It may be difficult to constrain any of these parameters individually by observing the variations.
\item The visibilities show silicate features in both type 1 and type 2 orientations. For type 1 AGN the appearance in ``absorption'' or ``emission'' has some strong dependence on $a$ and the feature is mostly canceled out around $a\sim-1$. However, the actual strength of the feature is degenerate with several parameters as in the case of SEDs (see Sects.~\ref{sec:specvis} \& \ref{sec:intf_sum}). Since the silicate feature in the visibilities is the result of both changes of flux and emission region size, it cannot be directly associated with emission properties of or within a certain region.
\end{list}

The presented models can be used as direct input for interpreting observational data. We are pleased to provide model SEDs and images based on our clumpy torus model (``\textit{CAT3D}'') to interested colleagues on request and via the webpage {\tt http://cat3d.sungrazer.org}. 

\begin{acknowledgements}
We thank R. Antonucci, P. Gandhi, K. Ohnaka, and K. Tristram for fruitful discussion and helpful suggestions during evolution of this project. The paper has been supported by Deutsche Forschungsgemeinschaft (DFG) in the framework of a research fellowship (``Auslandsstipendium'') for SH. We would also like to thank the anonymous referee who provided very constructive comments from which the manuscript substantially profited.
\end{acknowledgements}

\begin{appendix}

\section{AGN radiation}\label{sec:agn_radfield}

As AGN/accretion disk input spectrum, we use a four-component power-law which is approximately in agreement with the range of power-law fits to QSO composite spectra \citep[e.g.][]{Zhe97,Man98,Van01,Sco04}:
\begin{equation}
\Fnu \propto \left\{\begin{array}{lcl}
                         \nu^2 & & \nu < 10^{14}\,\Hz\,\,\leftrightarrow\,\,>3\,\micron \\ \nonumber
                         \nu^{1/3} & & 10^{14}-10^{15}\,\Hz\,\,\leftrightarrow\,\,0.3-3\,\micron \\ \nonumber
                         \nu^{-1} & & 10^{15}-10^{16}\,\Hz\,\,\leftrightarrow\,\,0.03-0.3\,\micron \\ \nonumber
                         \nu^{-2} & & > 10^{16}\,\Hz\,\,\leftrightarrow\,\,<0.03\,\micron\,\, (>0.4\,\mathrm{keV}) \nonumber
                      \end{array} \right.
\end{equation}
The main difference to the spectrum used in \citet{Hon06} concerns the relatively blue part with $\Fnu \propto \nu^{1/3}$ in the optical to near-IR between 0.3 and 3\,$\micron$. This spectral shape has recently been confirmed by near-IR polarimetry of quasars \citep{Kis08}. 

We note that the actual shape of the spectrum is much less important for the heating of the dust than the total (spectrally-integrated) luminosity of the accretion disk, $\Lbol = \int\Lnu\dd\nu$. In the optically-thick case, the total absorbed heating power is $L_\mathrm{abs} = \int Q_\mathrm{abs}(\nu)\Lnu\dd\nu \sim 0.5\times\Lbol$, depending on the actual dust composition, where $Q_\mathrm{abs}(\nu)$ is the absorption efficiency of the dust. Thus, the spectral shape of the incident radiation is integrated out in the process of radiative transfer.

\section{The influence of the accretion disk on IR interferometry of dust tori}

The presented torus models do not include any accretion disk contribution. For type 2 AGN, it is safe to conclude that most of the observed IR flux is coming from the dust torus since, by definition, the accretion disk has to be obscured. On the other hand, depending on wavelength, the accretion disk might contribute to the overall AGN IR flux in type 1 AGN. 

Recently, we presented an analysis of the near-IR emission of nearby type 1 AGN \citep{Kis07}. We used a color-color diagram to disentangle the hot dust emission from the accretion disk power law spectrum extending from optical wavelengths into the IR. A simultaneous power-law+black-body fit suggested that the accretion disk contribution to the AGN near-IR is around 10$-$50\% in the $H$-band and only 5$-$25\% in the $K$-band. The decrease of accretion disk contribution with wavelength is consistent with the recent observation of the blue tail of the near-IR accretion disk emission \citep{Kis08} which suggests that for any wavelength longer than $2\,\micron$, the accretion disk has marginal effect on SED and interferometry. In \citet{Kis07}, we also showed that a possible 20\% accretion disk contribution in the $K$-band increases the visibilities by at most $\la$0.1 for the longest baseline shown in Fig~\ref{fig:tor_vis_spec}. Since the position-angle variations in the near-IR are of the same magnitude, any accretion disk contribution will be hidden under these variations. Thus, we refrained from adding the accretion disk contribution to the presented figures and note that for interpretation of near-IR interferometry, a case-to-case analysis based on the color-color diagram should constrain to what degree the accretion disk has to be taken into account.

\end{appendix}

\bibliographystyle{aa}

\end{document}